\documentclass[acmsmall]{acmart}

\usepackage{xspace,color,soul, tabularx}
\newcommand{\ie}{{\emph{i.e.}},\xspace}
\newcommand{\eg}{{\emph{e.g.}},\xspace}

\newcommand{\etal}{{\emph{et al.}}\xspace}

\definecolor{lightblue}{rgb}{.93,.96,1} \sethlcolor{lightblue}
\newcommand{\pquote}[3] {{\textit{``#1'' (#3)}}}

\newcommand{\blue}[1]{{\color{black}{#1}}}
\newcommand{\red}[1]{{\color{black}{#1}}}

\AtBeginDocument{%
  \providecommand\BibTeX{{%
    \normalfont B\kern-0.5em{\scshape i\kern-0.25em b}\kern-0.8em\TeX}}}

\copyrightyear{2024}
\acmYear{2024}
\acmConference[CSCW '24]{ACM Conference on Computer-Supported Cooperative Work}{November 09--15, 2024}{San Jos\'e, Costa Rica}
\acmJournal{PACMHCI}
\acmYear{2024} 
\acmMonth{11} \acmPrice{} 
\usepackage{color-edits}
\addauthor[Nik]{nm}{black}
\addauthor[NikM]{nam}{red}

\begin{document}

\title[Power and Play in AI Ethics Discussions]{Power and Play:\\\smaller{Investigating ``License to Critique'' in Teams' AI Ethics Discussions}}
% \shorttitle{}

% \affiliation{%

\author{David Gray Widder}
\affiliation{%
  \institution{Digital Life Initiative, Cornell Tech}
  % \streetaddress{New }
  \city{New York City}
  \country{USA}}
\email{david.g.widder@gmail.com}

\author{Laura Dabbish}
\affiliation{%
  \institution{Carnegie Mellon University}
  % \streetaddress{New }
  \city{Pittsburgh}
  \country{USA}}
% \email{david.g.widder@gmail.org}

\author{James Herbsleb}
\affiliation{%
  \institution{Carnegie Mellon University}
  % \streetaddress{New }
  \city{Pittsburgh}
  \country{USA}}

  \author{Nikolas Martelaro}
\affiliation{%
  \institution{Carnegie Mellon University}
  % \streetaddress{New }
  \city{Pittsburgh}
  \country{USA}}
% \email{david.g.widder@gmail.org}

% \affiliation{%

% \affiliation{%

% \affiliation{%

% \affiliation{%

% \affiliation{%

% \affiliation{%

% \renewcommand{\shortauthors}{Trovato and Tobin, \etal}

\begin{abstract}

Past work has sought to design AI ethics interventions--such as checklists or toolkits--to help practitioners design more ethical AI systems. 
However, other work demonstrates how these interventions may instead serve to \textit{limit} critique to that addressed within the intervention, while rendering broader concerns illegitimate. 
In this paper, drawing on work examining how standards enact discursive closure and how power relations affect whether and how people raise critique, we recruit three corporate teams, and one activist team, each with prior context working with one another, to play a game designed to trigger broad discussion around AI ethics. 
We use this as a point of contrast to trigger reflection on their teams' past discussions, examining factors which may affect their ``license to critique'' in AI ethics discussions. 
We then report on how particular affordances of this game may influence discussion, \blue{and find that the hypothetical context created in the game is unlikely to be a viable mechanism for real world change.} 
We discuss how power dynamics within a group and notions of ``scope'' affect whether people may be willing to raise critique in AI ethics discussions, and discuss our finding that games \blue{are unlikely to enable direct changes to products or practice}, but may be more likely to allow members to find critically-aligned allies for future collective action. 
\end{abstract}



% \keywords{datasets, neural networks, gaze detection, text tagging}

\received{20 February 2007}
\received[revised]{12 March 2009}
\received[accepted]{5 June 2009}


\maketitle

\section{Introduction and Related Work}
Many companies have Responsible AI guidelines, which often revolve around principles like Fairness, Accountability, and Transparency~\cite{jobin2019global}.
However, research \blue{finds that structural and systemic limits encoded in Silicon Valley logics,} such as technosolutionism and the normalization of failure, limit those who seek to enact change~\cite{metcalf2019owning}.
High-profile incidents illustrate these limits. 
In separate incidents, Meredith Whittaker~\cite{nyt2019retaliation} and Timnit Gebru~\cite{metz2020google} departed Google after seeking to organize against building military drone tech and the company's handling of sexual harassment, and bias and environmental impacts of ever-larger language models, respectively. 
Such cases demonstrate that there are limits to the kind of direct ethical critique acceptable within tech companies. 

% In a more recent case, so-called ``godfather of AI'' Geoffrey Hinton left Google in the spring of 2023, ``so he can freely speak out about the risks of AI''~\cite{metz2023godfather} and not impact his employer while doing so.
%\footnote{https://twitter.com/geoffreyhinton/status/1652993570721210372}

The prolific adoption of Responsible AI standards in companies~\cite{jobin2019global} may initially seem to legitimize workers as they raise ethical concerns~\cite{madaio2020co}.
However, other work shows that they may, in fact, have the side effect of setting specific \blue{boundaries} on what is and is not legitimate to raise as an ethical concern.
In their work examining environmental sustainability standards in companies, Christensen \etal show how such standards enact ``discursive closure,'' which they understand in light of Deetz~\cite{deetz1992democracy} to mean legitimizing certain narrow kinds of employee critique while tacitly ruling others out of scope~\cite{christensen2017license}. 
They suggest that a ``license to critique'' must be deliberately created to work against this limiting of discourse so that such standards can be flexible and enable discussion of concerns they do not specifically enumerate.

Analogous arguments exist regarding the effect of responsible AI standards. 
For example, Keyes \etal satirically argue that a system can be Fair, Accountable, and Transparent yet still \blue{``turn the elderly into high-nutrient slurry,''} showing how \blue{certain} harms may be outside of the scope of discursively closed principles~\cite{keyes2019mulching}.
\blue{Greene \etal show that the language of Responsible AI standards focuses scrutiny on the \textit{design} of an AI system, but this focus thereby ``reject[s] critiques of business practice''~\cite{greene2019better}.} 
In this sense, principles \blue{such as Fair, Accountable, or Transparent that define responsible AI standards, or their focus on AI design rather than underlying business logics may set the discursive limits of the \textit{de facto} language of AI ethics, thereby rendering concerns outside of this language less legitimate or intelligible,}\footnote{\blue{Indeed, this constrained language of AI ethics may operate similarly to the \textit{de facto} language of ``doing diversity,'' in a way that similarly ``conceals inequalities and neutralises histories of antagonism and struggle.''~\cite{ahmed2006doing} }} both for workers in companies and for activists and policymakers seeking to influence companies' actions.
\red{Put simply, if an ethical issue is not addressed by a guideline or standard, then it may be seen as out of bounds for critique, thereby limiting which issues can be discussed. In particular, standards and guidelines often seek to help practitioners \textit{build better} AI systems~\cite{greene2019better}, instead of helping them to critique the (business) ends to which they may be put, limiting critique to technical implementation instead of questioning if the system should be built at all.}

\blue{While past research shows that information sharing among team members predicts outcomes of effectiveness such as solution correctness, profitability or market growth~\cite{mesmer-magnus2009information}, and highlights the complexity of team dynamics through analysis of car manufacturing environments~\cite{benders1999teams}, \red{some studies have examined} team discussions of AI ethics, a context where it may be less clear what constitutes a ``correct'' answer, or where different actors may evaluate the ethics of different outcomes differently.
For example, prior research has examined how teams implement AI ethics guidelines (\ie~\cite{jobin2019global}) into organizational processes~\cite{rakova2020responsible}, uncovering structural and systemic limits such as role uncertainty \red{where practitioners are unsure about their remit, and misaligned incentives between teams.
To overcome these issues, some have explored how workers can try to steer conversations within their organizations.}}
For example, Madaio \etal~\cite{madaio2020co} examine checklists as a way to enable team discussion on designing fair AI systems. 
Their study found that some workers were concerned that advocating for AI fairness issues \red{outright} may impact their career advancement or lead to them being ``labeled a troublemaker'', but found that a checklist may be able to ``empower individual advocates'' to raise issues that are legitimized by the checklist~\cite{madaio2020co}.
Related work on UX professionals seeking to steer their company's values \blue{show how they encounter similar limits, and thus resort to} ``soft'' tactics \blue{to attempt to} change values \blue{and practices} while still operating within perceived acceptable bounds \blue{within} their company \cite{wong2021tactics}.
\blue{Other structural and systemic limits include insufficient power within one's role to address ethical concerns~\cite{widder2023power,rakova2020responsible}.}
Furthermore, many AI ethics tools or guidelines are often focused on a subset of technical machine learning topics \cite{wong2022seeing}---such examples from CSCW include systems for leveraging crowds to develop ethical constraint specification of AI systems \cite{mandel2020crowd} or observing how datasheets for datasets may support ethical thinking \cite{boyd2021datasheets}.
\red{Such systems can improve the ethical implementation of AI and workers' abilities to engage with ethical AI considerations, at least within the technical areas such tools aim to address.}
%found that having an AI fairness checklist may be able to ``empower individual advocates'', by ``minimiz[ing] social costs'' incurred when one ``raise[s] concerns''~\cite{madaio2020co}. 

However, given the concerns about the discursive closure effects of standards and sets of principles outlined above \cite{christensen2017license}, we question whether enumerated lists of principles, checklists, or other tools rooted directly in technical aspects of machine learning can support workers in raising broad and varying ethical concerns. 
Additionally, the broader culture in which they are enacted---organizations with certain notions of ``efficiency'',  technological solutionism, and status hierarchies based on technical merit~\cite{metcalf2019owning}---may limit what concerns workers feel able to raise under these kinds of interventions. 
Checklists and other artifacts that seek to operationalize AI ethics standards may enact discursive closure, limiting discussion to those issues they enumerate rather than enabling a broader license to critique.
Prior work analyzing AI ethics toolkits find that they often frame the work of AI ethics to be narrow technical work, rarely engaging with wider social issues or the power dynamics in which this work must take place~\cite{wong2022seeing}.

To this end, we ask the research question: \textbf{RQ1: What factors appear to influence members' ``license to critique'' when discussing AI ethics with their team?}
While many have interviewed AI practitioners individually about ethics issues and processes~\cite{veale2018fairness, holstein2019improving, madaio2020co, widder2022limits,Varanasi2023hodgepodge}, group dynamics influence how discussion proceeds. 
We have only identified one study which examines \textit{group} discussions of AI ethics, however, this study was \textit{a priori} scoped to Fairness~\cite{madaio2022assessing}, foreclosing discussion of wider concerns as discussed above.
Understandably, answering questions about group discussions through direct observation is difficult to study---often, by their nature, AI ethics conversations in companies involve proprietary information and ethical or legal issues that may be highly sensitive. 
\blue{Past ethnographic research examines similarly sensitive questions in public sector and medical contexts, for example, the causes of the Challenger disaster at NASA~\cite{vaughan1996challenger}, trust in AI tool use for space mission software at NASA ~\cite{widder2021trust}, secret nuclear weapons development at a National Laboratory~\cite{gusterson1996nuclear}, and cultural barriers in efforts to shorten doctors' shifts in hospitals to improve patient safety~\cite{kellogg2009operating}. This work grapples with and successfully negotiates questions of researcher access in order to enable long-term study of similar questions. However, these may be difficult to transfer to for-profit companies: public sector institutions often have mechanisms for public oversight and scrutiny, and medical contexts have existing professional norms, fiduciary duties, and legal accountability that do not exist in AI companies~\cite{mittelstadt2019principles}.}
Even in other contexts, such as in activist groups, discussions may be hard to observe as they are often unplanned, spontaneous, or involve sensitive plans that the group may be unwilling to reveal.
\red{Anecdotally, many teams we approached were reluctant to grant us access, citing the sensitive nature of their work.}
%, and how norms which inscribe discursive closure may affect discussion -- and perhaps understandably so: these discussions likely involve sensitive legal and ethical matters, as well as intellectual property and trade secrets, making participants understandably unwilling to have these interactions studied, recorded, or listened to.

To overcome these challenges, we asked existing teams who have experience discussing AI ethics in their team to discuss AI ethics scenarios in a \red{\textit{hypothetical context}} created for our study (described below). 
% This is done to minimize the risk of revealing sensitive information.
We recruited three teams across two companies and one activist group.
Recruiting real teams allows them to bring their associated shared experiences and context, shared understanding of process, and existing power dynamics into the study discussions. 
Including an activist group provides a point of comparison from which to question norms in company contexts, \blue{so as to imagine other possible practices}.
In individual follow-up interviews, we used participants' experience discussing AI ethics in this hypothetical context as a probe to enable them to reflect on differences and similarities between it and AI ethics discussions in their ordinary team context. 
In short, we set out to learn about how organizational and team norms influence discussion of AI ethics, by using a hypothetical context to (a) enable participants to speak more freely in contrast to sensitive company discussions, and (b) serve as a probe that participants can compare to their past experience. 

In designing a hypothetical context to facilitate AI ethics discussions, we also sought to study factors that may help create a license to critique within these discussions. 
In his book \textit{Domination and the Arts of Resistance}, James Scott drew from his fieldwork to argue that people speak and act differently depending on power differentials between them and their audience, with less powerful subjects using ``public transcripts'' when in earshot of the powerful, while persistently using ``hidden transcripts'' when speaking ``offstage ... outside the intimidating gaze of power''~\cite{scott1990domination}. 
Scott emphasizes continuity between these two stages, in particular, that ``rumors, gossip, folktales, songs, gestures, jokes'' are where people may dissent more freely while ``hiding behind anonymity or behind innocuous understandings''~\cite{scott1990domination}.  

Motivated by Scott's concept of offstage talk~\cite{scott1990domination}, we see a connection with games based around speculative futures as a way to provide an ``innocuous'' context for discussion.
\red{In this work, we explore how a game can be used as a method to study conversation around AI ethics and to shape conversations between team members.}

We look toward the literature on speculative futures and the power of speculative games to create more playful contexts which may help resist discursive closure. 
In their book \textit{Speculative Everything}, Dunne and Rabby articulate how using speculative futures exercises can allow teams to ``explor[e] alternative scenarios'' to enable them to ``be discussed, debated, and used to collectively define a preferable future''~\cite{dunne2013speculative}. 
Mankoff \etal articulate the value and methods of Futures Studies within human-computer interaction, and in particular the value of ``critical reflection'' to examine ``the relationship between present-day realities and potential futures'', mentioning the possibility of fiction and multiplayer games to support this critical reflection~\cite{mankoff2013looking}.
In a technology context specifically, \textit{Project Amelia} used immersive theater to encourage participants to reflect on their privacy behavior within technology~\cite{skirpan2022privacy}.
\blue{Past research reviews and draws together Speculative Design with Games Studies approaches, concluding that ``explorative worlds produced by games [have] much to offer over the traditional mediums of speculative design due to their inherent interactivity''~\cite{coulton2016games}, and a past meta-analysis demonstrates how games have been used as data collection methods~\cite{smith2015metaanalysis}. 
Flanagan and Nissenbaum have also shown how digital games embed values, and may be used in 
``animating personal, political, and artistic expression''~\cite{flanagan2014values}.

There are also at least two existing examples of past works
proposing games which employ speculative approaches in an AI ethics context.}
Ballard \etal's \textit{Judgment Call} was designed around Microsoft's articulated ethical principles
to create ``space for difficult or uncomfortable conversations'' ~\cite{ballard2019judgment} \blue{and argued for the applicability of design fiction methods in game contexts}. 
Martelaro \& Ju's \textit{What Could Go Wrong?} is a game where participants combine cards outlining a particular scenario with cards naming a particular user group or exceptional circumstance, and discuss concerns emerging from these new combinations. 
To examine the possibility for these kinds of games to create the ``innocuous'' understandings that Scott wrote of \cite{scott1990domination}, and to examine how they may work against discursive closure~\cite{christensen2017license}, we pose another research question:
\textbf{RQ2: How do AI ethics discussions unfold while playing a game oriented toward speculative critique?}

% \nmedit{In addition to understanding how a speculative game might shape conversation among a group, we are also interested in understanding if conversations in a speculative game may be different conversations while using an AI checklist.
% This raises our third research question: \textbf{RQ3: How do conversations held during gameplay compare and contrast to conversations facilitated by a checklist?}
% To answer this question, we have a group of 29 students in a responsible AI course holding discussions facilitated by either the game or an AI checklist.
% We collect responses from a post-class questionnaire to compare the kinds of conversations using each intervention.}

We next describe how we observe four existing corporate and activist teams as they play the \textit{What Could Go Wrong?} game, and conduct one-on-one follow-up interviews to compare and contrast their conversations during the game with their perceptions of their past typical discussion of AI ethics.
We use the game to provide a point of comparison for participants, to allow them to more easily reflect on their ordinary discussions of AI ethics, 
and how they may or may not feel license to critique (RQ1).
We find that notions of ``scope'' bound the kinds of concerns that can be raised in AI ethics discussions, and how this is inflected by group power dynamics. 
We then look at the specifics of the conversations in the game (RQ2).
We find that a game context can broaden conversation, but that games may be unlikely to lead to change directly \blue{due to existing power dynamics and the disconnect between low-stakes hypotheticals and higher stakes changes to the product.} \red{However, we also see that games} may help teammates better understand each others' critical orientation and thus may help form collectives for future action. 
Our results help AI ethics research better account for team power dynamics, and have implications for research where games are framed as interventions~\cite{wilkinson2016brief}.
\blue{Our study contributes a critical evaluation of the limits and opportunities for using games as a tool to help professional teams discuss sensitive issues like AI ethics (RQ2), and uses this tool as a probe in data collection to investigate factors that influence team members' ``license to critique'' when discussing such issues in their ordinary work (RQ1).}

\section{Methods}
We engage three teams from companies and one group of activists to first play the \textit{What Could Go Wrong?} game, and then follow up with one-on-one interviews to probe on differences between conversations had during the game and their past AI ethics discussions. 
% After completing the sessions with the company and activist teams, we answer RQ3 comparing conversations facilitated by a speculative game and a responsible AI checklist by engaging 29 students in a responsible AI course, with half playing the game and half using the checklist.
% We describe the game and AI checklist interventions below.

\subsection{Procedure}
Given that we seek to examine how games affect AI ethics discussions rather than build a game ourselves, we choose to conduct the present study using Martelaro \& Ju's \textit{What Could Go Wrong?}
a game in which groups of 4--5 participants discuss a series of AI applications and potential harms.
Other games for AI ethics discussion exist~\cite{ballard2019judgment}, but we choose this one because its source materials are readily available, can be easily adapted by adding cards, includes an online version for remote participants,\footnote{\url{https://github.com/nikmart/what-could-go-wrong-ai}} and because it is modeled on popular party games \textit{Apples to Apples} or \textit{Cards Against Humanity}, which may make it more easily understood by participants.

In this game, players first select a random \textit{Prompt} of a particular automated technology (\eg ``autonomous food delivery'') and then each chooses one \textit{Response} card which includes particular user groups (\eg``blind user''), events (\eg``non-consenter infringement'', ``random crashes'') or exceptional circumstances (\eg``locusts'') to create scenarios in which to discuss the  game's eponymous question.
A ``Card Czar'', who does not play a \textit{Response} card, leads the discussion by either choosing one \textit{Response} to match with the \textit{Prompt} or by discussing all cards played.
The round ends when the group decides to move forward and the Card Czar chooses one card as the winning combination.

\red{In our study,} gameplay sessions lasted 1.5 hours.
\red{Participants played through many rounds with different \textit{Prompts}.}
There was no time limit set for each round, with groups completing between three to eight game rounds each.
% The original What Could Go Wrong? was developed specifically for Autonomous Vehicle applications and was released online~\cite{martelaro2020could}.
% We built upon the original game and expanded the set of \textit{Prompt} cards to include a variety of AI applications spanning topics such as autonomous vehicles, robotics, financial technology, and generative AI.
% We also increased the number of \textit{Response} cards from the original set.

% Teams played either online, using PlayingCards.io\footnote{www.playingcards.io} (a web-based game simulator) and video conferencing, or using a physical deck of cards while seated around a table.
% The online version of our AI version of the card game and the full listing of \textit{Prompts} and \textit{Responses} is made available under a Creative Commons license\footnote{[Github to game source, removed for submission]}.

% \nmedit{\subsection{AI Checklist}
% In the session comparing the card game with an AI checklist intervention, student groups working with the checklist were provided a printed sheet containing the same prompts as were printed on the prompt cards, but instead of using the game intervention, they examined the prompted scenarios using a Responsible AI checklist, namely Microsoft's AI Fairness Checklist as it is an exemplar checklist and has been the subject of past research.
% The session lasted 1.5 hours.
% }

\subsection{Participants}
% \usepackage{booktabs}
\begin{table*}[ht]
\begin{tabular}{llllll}
\toprule
% \multicolumn{2}{c}{Country where...} \\
Resides & Citizen & Gender & Current Role & Yrs in org & Highest Degree, Field \\
\midrule
\multicolumn{6}{c}{Company 1, Team 1 \textit{(C1-T1) } ---  Remote Session} \\
\midrule
USA & USA & Woman & Research Scientist & <1 yr & PhD, Social Sciences \\ %5 months,  Communications 
USA & USA & Man & Research Scientist & >25 yrs & PhD, Social Sciences \\ %, 26, Psychology
Germany & India & Man & Research Scientist & 1-5 yrs & PhD, Computer Sciences \\ %1 year & PhD, Artificial Intelligence \\
Mexico & Mexico & Man & Research Engineer & 1-5 yrs & MS,  Computer Sciences \\ %1.5 years & MS, Artificial Intelligence \\
USA & India & Man & Research Scientist & 1-5 yrs & PhD, Computer Sciences \\ % 3 years Computer SCeince 
\midrule
\multicolumn{6}{c}{Company 1, Team 2 \textit{(C1-T2)}  ---  Remote Session} \\
\midrule
USA & USA & Woman & Program Manager & 5-15 yrs & MS, Humanities \\ %11 years & MS, International Development \\
USA & USA & Woman & Program Manager & >25 yrs & MBA  \\ %26 years & MBA  \\
USA & USA & Woman & Director & 15-25 yrs & MS, Computer Sciences \\ % 20 years & MS, Computer Engineering \\
\midrule
\multicolumn{6}{c}{Company 2, Team 1 \textit{(C2-T1)}  ---  Remote Session} \\
\midrule
USA & USA & N.B. Femme & Research Scientist & <1 yr & PhD, Social Sciences \\ % 6 months & PhD, Computational Social Sci. \\
USA & USA & Woman & Research Scientist & 1-5 yrs & PhD, Computer Sciences \\ % 1 year & PhD, Computer Science \\
USA & USA & Woman & Research Scientist & 1-5 yrs & MA, Humanities, Social Sci. \\ % 3 years & MA, English linguistics \\
USA & USA & Woman & Research Engineer & 1-5 yrs & MS, Computer Sciences \\ %3 years & MS, Computer Science \\
\midrule
\multicolumn{6}{c}{Activist Collective \textit{(AC)}  ---  In-person session} \\
\midrule
USA & India & Woman & Masters Student & 1-5 yrs & BA, Arts \\ %& 1.5 years & BFA \\ Masters of Fine Arts student
USA & USA & Woman & PhD Student & 1-5 yrs  & BA, Humanities, Arts \\ %2.5 years & BA, Bioethics \& Design \\
USA & Russia & Woman & PhD Student & 1-5 yrs  & MA, Social Sciences \\ %4 years & MA, Sociology \& Political Sci. \\
USA  & USA & Woman & Designer & <1 yr & MS, Arts \\ %6 weeks & MS, Urban Design \\

\bottomrule
\end{tabular}
\caption{Participant demographics.}
\label{tab:interview-participants}
\end{table*}

We recruited 16 participants across four teams, all who had prior history of discussing ethical issues in AI, existing norms of interaction, and a shared organizational context.
%spanning two companies with existing responsible AI practices similar to most major technology companies~\cite{jobin2019global} and one anti-carceral tech activist group. 
% Since we are interested in investigating how existing organizational and social norms affect AI ethics discussion, and how existing teams with shared context (particular norms, hierarchies, working relationships and shared knowledge) engage a game-based AI ethics intervention, 
The first two teams were from a US-based multinational technology company; the first team works on research and engineering for an AI-enabled hardware deployment designed to observe and aid workers in manufacturing environments, and the second provides ethical evaluation and guidance for products and services the company develops. 
The third team worked at a European-based multinational media streaming company, all of whom conduct research and development work to build algorithmic features and evaluate them for possible ethical issues.
The fourth team included members of an activist collective focusing on raising awareness of carceral technology developed and deployed in the city where they are based.
Their participation provided a contrasting set of team norms, less influenced by strict hierarchies or tech company practices.

Company teams played the game over a video conferencing platform using an online card table simulator\footnote{www.playingcards.io}, and the activist group played in person using a printed deck, reflecting how each of these teams ordinarily met together. 
\blue{We note that past work shows that virtual versus in-person environments affect how teams discuss creative ideas~\cite{brucks2022virtual}, and we discuss how affordances from virtual environments affected discussions in Sections~\ref{sec:push} and \ref{sec:vulnerability}.}

\blue{To protect anonymity and reduce re-identification risk, some demographic descriptors here have been generalized in Table~\ref{tab:interview-participants}, and we provide group rather than individual identifiers alongside quotes, even where quotes originated from individual follow-up interviews: \ie C2-T1 refers to a quote from a participant at Company 2 Team 1, and AC refers to a participant in the Activist Collective. Where aspects of a participant's identity are important for contextualizing what they say (\ie how one's gender affects their ability to raise critique), we selectively note this context adjacent to their quotes, thereby attempting to balance context with anonymity.}
While team C1-T2 played with their manager (which we discuss in Section~\ref{sec:managers}), no other company teams did, but we note significant diversity in team seniority (which we discuss in Section~\ref{personal-atributes}) as can be seen in Table~\ref{tab:interview-participants}.

% Both companies had responsible AI programs, , which in these cases included things like reading groups which convene to read and discuss research on issues like algorithmic fairness, ethical review committees which interface with product teams, and dedicated research into related issues relevant to the companies' products.

% \subsubsection{Comparative session with students}
% \nmedit{The session comparing the game to the AI Fairness checklist involved 29 students taking a one-semester computing ethics and policy course. 
% Student participants did not have prior context working together, with many not knowing each other outside of the classroom. 
% All opted to participate in a follow-up survey, and 28 opted to provide demographic information: 18 men and 10 women; the majority were undergraduates with a few masters students, studying subjects ranging across Computer Science, AI, Math, Physics, Civil Engineering, Business, Public Policy, and Architecture, and with previous work or internship experience spanning Software Engineering, ML Engineering, Research, Data Science and Technical Writing. 
% 19 students were from the USA, four from China, and the rest from Australia, South Korea, and Turkey.
% %https://docs.google.com/spreadsheets/d/1v0pt1nR4Q1Bi7ui4Gt-O_mS_3JF61czv-kJqswLgsng/edit#gid=1012160450}
% }

\subsection{Data collection}

Each team played the game in a session lasting 1.5 hrs.
The conversations for each team were audio recorded with consent and IRB approval. 
Except for one participant who declined, all participants participated in a one-on-one semi-structured~\cite{weiss1995learning} recorded follow-up interview (lasting an average of 34 minutes, but as long as 49 minutes or as short as 22 minutes). 
\red{During the interview, we asked participants} to reflect on in-game experiences and conversations which arose, but with a special focus on contrasts between in-game discussions and past AI-ethics discussions within their team or organizational context.
We also focused on the extent to which they did or did not feel comfortable raising anything or disagreeing, during in-game or prior AI ethics discussions. 
For remote sessions, researchers turned off their cameras unless answering questions to minimize any effect their presence may have had on the teams' discussion.

\subsection{Data analysis}
We analyzed data under an interpretive epistemological paradigm~\cite{lincoln2011paradigmatic} using thematic analysis~\cite{clarke2021thematic}.
\blue{The first author} began by open coding~\cite{strauss1990basics} a selection of six transcripts of two play sessions and follow-up interviews selected for diversity of role, company, team, and past experience, annotating portions relevant to our research questions, while collating coded portions and associated themes into an analysis document.
After reoccurring themes emerged in this document, a tentative code book~\cite{clarke2021thematic} comprising five initial codes was constructed, and applied to all transcripts. 
In approximately weekly meetings between authors, new codes arose to capture new themes of interest, in which case the coding frame was updated and data re-coded in an iterative process.
Two categories---those addressing our research questions specifically---had a large amount of divergent data, so coded portions were printed out and sorted into finer grained cohesive categories using an open qualitative card sort~\cite{zimmermann2016card}.
\blue{The nature of our data collection and interpretive epistemological stance yields rich insight into participants' experiences and how they make sense of them, but results should not be directly generalized into new contexts without further study.}

% Given that the focus of the present study is on how organizational norms and existing social ties affect AI ethics discussion, data from existing teams' sessions and their individual follow-up interviews will comprise the focus of the findings we report below. 
% However, we also present survey data from student participants when it contrasts with data from existing teams, as appropriate. This data was analyzed after the data from sessions and interviews with existing teams, using the same coding frame as a starting point, but also annotating other present differences and similarities.
% We present the results of our analysis in response to our three research questions.

% Maybe put this quote into ``vs checklists'' section, as a complementary to checklists: 
% \pquote{as a result, we also didn't really discuss some of the obvious things. }{L}{C1-T2}
% yet:
% \pquote{I think the things that are not necessarily the most obvious are the ones sometimes that spur interesting conversations. }{C1-T2}{}

\section{RQ1: What factors influence members' ``license to critique'' when discussing AI ethics with their team?}
\blue{In this section, we discuss factors that influence team members' license to critique when discussing AI ethics.
To set the stage, we first show how techno-optimistic organizational norms mean that critique is often perceived as ``too negative'', after which we narrow in on two particular ways that this occurs.
Firstly, license to critique is regulated through a notion of ``scope'', bounding the issues a team believes it can or will take on, enacted pragmatically through time pressures which curtail discussion of ``edge case[s]'' and to issues solvable through ``technical solution[s]'', reified by role divisions which compartmentalize ethical concerns. 
Secondly, we show how ``who is in the room?'' matters for resulting ethics discussions, and is affected by whether one's manager is present, one's understanding of teammates' orientation toward critical topics, and the ways that discussants' personal identity (\eg age, gender, experience) affects who is credibly able to raise concerns.} 
\blue{In this section} we rely predominantly on data collected during one-on-one follow-up interviews, \blue{with participants' experiences during the game secondarily serving as a probe to enable reflection \red{and contrasts} to ordinary discussions of ethics within their organizations}.

% Here, the goal is that participants may have felt free by virtue of game or research context, and either is fine with us. Both provided a contrast to participants ordinary manner of discussing AI ethics -- and thus provided a contrast in experience to probe participants to reflect on their ordinary habituated experience. 

% In this section, we primarily focus on participant's reflections on their typical AI conversations based on their contrasting experience during the gameplay. 
% We now report findings for each of our two research questions, beginning with RQ1, which largely draws on follow-up interviews where participants reflected on their team's normal AI ethics discussion, using the game discussion as a probe to provide a point of contrast. In the next section, we present findings to RQ2, examining how game affordances related to in-game discussion, largely drawing on observation and recordings of in-game discussions.

\subsection{Organizational norms push against ethical critique}\label{sec:push}
Across company participants, a feeling of organizational norms, often implicitly understood, modulated whether they feel able to bring up ethical issues during their typical conversations around AI in their work.
For example, one participant reflecting on past deliberation around AI ethics noted an expectation that there should be a \pquote{significant enough concern}{J}{C1-T2} before she would \pquote{have a conversation about it}{J}{C1-T2}, suggesting that raising issues should be saved for only the most dire cases.

Other participants relayed how it felt difficult to raise ethical critique about new technologies, in the face of wider company excitement \blue{and techno-optimism surrounding new technologies}. She said that \pquote{just saying no [<about a product idea, redacted>] just makes everybody frustrated [...] striking that balance is something I'm still learning how to do properly. And so it takes some work [and] conscious effort}{L}{C2-T1}. 
She gave the example that during ``large forum'' company meetings, when someone is presenting new technology that she might have ethical issues with, there is often \pquote{a lot of enthusiasm going in, [which] I think make[s] it hard to kind of speak out. [...] you've got like all these like, emojis like `thumbs up', `loving it', and then like the chat is blowing up with people saying how amazing [the tech] is}{D}{C2-T1}.

% He also displayed some awareness that he \pquote{tend[s] to speak a lot. [but now] I'm kind of more cognizant of where you know, like, I make sure I don't talk too much. [...] but it ends up happening [that I speak a lot].}{RM}{C1-T1}.}
Another participant \blue{who identified as a woman} discussed how she was self-conscious that negativity might go against company norms and forward progress for their team, relaying that she had been told that she is\pquote{too negative at work. And don't focus enough on the positives [...] It's difficult to pull that back}{E}{C1-T1}, but that she was recently\pquote{trying to be a `team player', and really trying to hold back when I disagree}{E}{C1-T1}. After being told by a close confidant that \pquote{you complain a lot [...] you shouldn't do that}{E}{C1-T1} she has\pquote{been trying to ramp back on disagreements and save it for when I feel most passionately}{E}{C1-T1}. \blue{Her particular concern about being told she is ``too negative at work'' alludes to past work examining how women are viewed as themselves a problem when they speak up about problems at work~\cite{ahmed2021complaint}. We further discuss relationships between identity and license to critique in Sec.~\ref{personal-atributes}.}

\blue{
\subsubsection{Summary}
In this opening subsection we have seen how, broadly, organizational norms enact discursive closure~\cite{christensen2017license}  whether through an implicitly understood need for concerns to be ``significant'' to merit discussion, as well as more quotidian worries about frustrating excited colleagues, or being perceived as too negative.}

% She said \pquote{it's been difficult to [...] balance [...] critique of a system with productive feedback on a system.}{E}{C1-T1}

 %Another participant reflected that his colleagues would often be \pquote{pretty conservative in issuing criticism. So [they] agree right away, like, even if someone is saying wrong things [they say] that's a good idea.}{RM}{C1-T1}, even agreeing with ``meaningless stuff'' that he says.

\subsection{``Scope'', its contestations, and its effects}\label{sec:scope}
\red{From our interviews,} we saw the notion of \textit{scope} invoked to set boundaries on what corporate teams believe they can or will take action on, and thus where discussion is focused. 
\red{Scope} appeared as a softer way to limit bounds of discussion---saying a statement is ``out of scope'' is not a judgment that it was incorrect or imaginary---but that it was beyond what a group believes organizational norms or incentives would permit them to discuss or take action on. 
We found that scope is often enforced through reference to time pressures, a push to solve particular problems often through technical means, and through role divisions leading to compartmentalization of ethical questions.  

\subsubsection{Scope: broadness or narrowness of critique}
Various notions of ``scope'' surfaced as key attributes defining what participants consider acceptable to raise in work discussions about AI ethics. 
For example, in contrast to game discussions, one participant said, \pquote{a lot of the discussions I've been having recently have been much more narrow}{L}{C2-T1}, often focused on specific aspects or features of a product. 
Another participant suggested that her work discussions about AI ethics don't have the \pquote{sense of freedom to go off and think about very unlikely harms that could happen and discuss those further}{K}{C1-T2}. She went on to say how she would like to integrate some of her outside ``passions'' into AI ethics discussions at work, for example ``dystopian'' themes from \pquote{watching shows like Black Mirror and reading all of these, you know, sci-fi dystopian stories}{K}{C1-T2}, but \pquote{those are things that I probably wouldn't share in an Ethical Impact Assessment review}{K}{C1-T2}, because it wouldn't be ``applicable.'' 

Participants perceived that certain kinds of harms or remedies, such as climate harms or those requiring ``diffuse'' systemic change can often be seen as out of scope and hard to discuss. 
For example, one participant relayed how someone brought up in a large team setting how
\pquote{LLMs have like a major climate impact. [...] And you try to bring these things up. [...] it's kind of falling on deaf ears that are unwilling [to hear this,] they're just kind of like, `nope, we're being told use LLMs, [so] we're using it.'}{D}{C2-T1}
Another suggested that \pquote{AI ethics is that it often falls short of trying to work towards systemic change}{R}{C2-T1}.
One participant spoke about how she feels most AI ethics conversations don't talk about more ``diffuse'' cultural effects, elaborating \pquote{Like what does it mean for people to use technology kind of pervasively in a specific way? [...] we don't, as often, I think, talk about [this]}{F}{AC}.
% David: pull quote where someone in ACT often spurred by news events
% David: Discuss how this person does not have a strict managerial heirarchy/ organizational closure to scope debate, but some group member described how they discuss events in the news, presenting a different kind of reactive closure, reacting to events to discuss or protest around, rather than proactive speculation of future. (somewhat similar to Christiantensen etal ``closure by the past'')

However, some ethical issues were seen as so potentially damaging to the product success, that they ceased to be ethical questions, instead expanding in importance to become ``product questions'', as one participant relayed: \pquote{big enough problems that [...] it's beyond the ethical questions. It's also just a product [question]}{L}{C2-T1}.
This suggests that something being an ``ethical question'' may be its own kind of limiting or minimizing scope.

\subsubsection{Time pressures and questions of relevance}
Participants reflected concerns in follow-up interviews about how time pressure scopes discussion only to ethical issues seen as most ``relevant''. 
Multiple participants reflected on time pressures within their teams, showing how this served to continually foreclose discussion of less direct yet pressing--in the eyes of participants--concerns. 
Others suggested how engineers on their broader team may perceive AI Ethics conversations as lacking relevance to their work. 
One said that while some engineers thoughtfully think through ethical issues, some engineers push back with questions of relevance: \pquote{sometimes the pushback of `oh, that's an edge case, that's never going to happen.'}{D}{C2-T1}. 
Others suggested that there might not  \pquote{be too much enthusiasm}{J}{C1-T2}, because it would be perceived as \pquote{a big chunk of time without a great ROI [return on investment]}{J}{C1-T2}, or that some might not \pquote{see how it would [...] be applicable to the work}{RB}{C1-T1} they do.

% : \pquote{maybe lack of time is the important word here}{S-chips}.
%David: make clear multiple participants said this. Common response. 

% When reflecting on whether they thought other teams would be willing to play the game to discuss responsible AI and potential harms, negative answers centered around time. 
% One stated that teams may be \pquote{crunched for time [...] people would have other projects, and they cannot make time}{S}{C1-T1} to play the game. Another said willingness would \pquote{depends a lot on their calendar.}{G}{C1-T1}.
% In this way, time becomes a factor limiting the scope of acceptable discussion. 
% \nmedit{Where time \textit{is} spent on ethical conversations reflects the discussion of issues seen as most valuable, with such conversations being linked to questions about the product, not just broader ethical implications.}

\subsubsection{Push to ``solve'': discursive closure by being scoped to ``fix'' a particular system}
In a similar sense to what Christensen \etal call ``closure by design'' in sustainability standards~\cite{christensen2017license}, participants discussed how in past AI ethics discussions, a ``goal orientation'' affected, and to an extent, limited, the kinds of conversations which arose. 
In some ways, this makes sense: participants in companies often were tasked with evaluating existing or proposed systems for ethical concerns, and then suggesting mitigations---largely technical changes to the system---to (partially) resolve those concerns. In other ways, such technosolutionist or technochauvinist thinking can entrench existing inequities, and obscure other ways of thinking or conceiving of problems or solutions~\cite{cunningham2023grounds}. 
\blue{This demonstrates how organizational dynamics and structure limits the kind of solutions seen as possible, by fracturing questions of accountability\cite{widder2023dislocated} between ``what to build'' questions and the narrower and technical scope of ``how to build'' that teams felt power over.}
Along these lines, some participants reflected on how the ``problem-solution'' script could preclude discussion of wider changes, such as systemic change, and does not fit into their prescribed role.
% or common ways of discussing ethical concerns about \textit{particular products} within their organization. 

Some talked about how AI Ethics conversations are often prompted by a specific product or problem: \pquote{maybe we're talking about large language models [and the] impacts of generation on, like, artists [...] So it's a little bit more targeted}{L}{C2-T1}, comparing this to in-game discussion which \pquote{felt more generative [because] there wasn't as much of a goal}{L}{C2-T1}.
A participant also noticed how conversations often \pquote{jump towards like, what's, what's a good technical solution?}{L}{C2-T1}, continuing to say that most past conversations were \pquote{solution-oriented, [like] if we're looking for a mitigation, what's the best or most practical way that [...] we can do that [...]you're trying to get at a smaller solution space}{L}{C2-T1}.
In this way, narrowing the scope of discussion was viewed as a process of ``solving the problem.''

Participants even spoke of how internal ethics tooling and processes lead to discursive closure: \pquote{the tools internally, they're a bit more guided [saying] `if you're interested in building a system or model, here are a bunch of questions that we want you to answer' [they] tend to be a lot more directed}{L}{C2-T1}, for example asking  narrower questions about whether a system she might work on uses protected demographic information. She mentioned, therefore, \pquote{They're a little bit less broad in terms of [...] societal impacts off [of our] platform. [...] the focus feels a little bit narrower}{L}{C2-T1}.
A participant from a different company spoke of this too, suggesting that conversations in-game were more ``creative'' than when dealing with ``reality'' when \pquote{having all the details, like we do [when we do] an Ethical Impact Assessment}{J}{C1-T2}

Participants suggested a variety of possible reasons for this rush to discuss technical solutions. One suggested that in a \pquote{tech company}{D}{C2-T1}, with an \pquote{engineering mindset}{D}{C2-T1}, open-ended introspection is \pquote{not always the vibe}{D}{C2-T1}. 
One participant reasoned that this may be due to \pquote{what is possible to change}{R}{C2-T1} within an individual worker or team's power, but also due to a cultural mindset biasing towards being \pquote{able to measure particular harms, the things that are not measurable, end up not being as easy to solve for}{R}{C2-T1}.

% And sometimes the solution is not technical in the first place, or the first direction is, it's something completely different. And it's nice to sort of take a step back and realize that and having [...] a structured way to do that, it seems quite nice. (l music, follow on quote)

\subsubsection{Role divisions leading to compartmentalization of ethics}

Some participants discussed how role divisions affect who is expected to care about, and handle, AI ethics questions. 
For example, one said that this wasn't his job, saying these issues were handled by a specialist committee, and a member of his team who \pquote{target[s] or address[es] those topics on [our] projects [...] we have people that do that}{G}{C1-T1}.
% \pquote{we have people that do that [...] on the team [...] cascade us [...] the information [...] }{G}{C1-T1}, and that he makes technical changes to his team's application accordingly. 
However, many others were concerned about this apparent ``compartmentalization'' of ethics: \pquote{the compartmentalization of what we do with any individual horizontal capability, I think this is a huge problem with respect to ethical uses of AI}{RB}{C1-T1}.
\blue{Ethics was seen as a function of one narrow specialist committee, and called out as an example of a broader phenomenon of the compartmentalization of horizontal capabilities, thus demonstrating how organizational dynamics narrow both who feels able to raise ethical critique, and how broadly or narrowly segmented this critique is able to be.}

Participants spoke about countering this compartmentalization, saying they \pquote{we need more diverse thoughts here}{RM}{C1-T1} and seek to \pquote{strengthen the bonds among some of the product, ML product practitioners, and me [in her ethics-focused role]}{R}{C2-T1}.
\blue{In this way, we see different ways that participants construct their role and and professional duties with respect to ethics---either as something that is best left to teammates assigned specifically to do this work, or conversely, as something that should not be compartmentalized, thereby still recognizing compartmentalization as a default practice bounding their own role, and hoping to work against this.}

\blue{
\subsubsection{Summary}
In this subsection we have shown how an often implicitly understood notion of ``scope'' mediates what kind of critique is rendered acceptable, thereby constituting that which is rhetorically ``closed'' in the enactment of discursive closure~\cite{christensen2017license}. We have further shown how scope is enacted and maintained in organizational and work practices, including how \textit{time pressures} lead to discursive closure due to perceived relevance,  a \textit{push to solve} issues focusing discussion on issues addressable within technical changes in the project at hand, and how \textit{role divisions} lead to a compartmentalization effect where some see ethics questions as not within the scope of their role.}

\subsection{``Who is in the room'': participants' power and critical orientation}

\subsubsection{If managers (or others with power) are in the room}\label{sec:managers}
More than just demonstrating a general awareness of who is in the room and how that affects what is safe to share, participants appeared acutely aware of and concerned with how bosses and managers shape the conversation. 
One noted that compared to other conversations, the game was a space where: \pquote{your boss isn't here}{E}{C1-T1} nor was the session being recorded by her employer.
Therefore, \pquote{you're free to talk about things that you think are weird or risky}{E}{C1-T1}.
Another participant said that in the play session \pquote{the things that I discuss here, it's not going to impact my paycheck next month. So it's more comfortable}{RM}{C1-T1}. 
One participant commented on the
\pquote{surveillance technology that's on everyone's [company] laptops}{R}{C2-T1}, 
and also on \pquote{worker exploitation}{R}{C2-T1} during the play session, but 
noted she wouldn't feel comfortable \pquote{bringing [this] up when it's not just around peers [and if] we had managers in the room [who are] on the company side}{R}{C2-T1}.

Some participants suggested that disavowal of their critique was sometimes justified by managers using an ostensibly altruistic rationale, saying \pquote{`I'm trying to make your life easier, we don't need to do this.'}{D}{C2-T1}.
Others reported their \pquote{manager, and like, my skip level [managers]}{K}{C2-T1} encouraging her to 
prioritize work that \pquote{they felt would be more impactful in a product}{D}{C2-T1}.
However, this did not always appear to be the case. In one play session including a manager, their subordinates said \pquote{I don't feel like there's really censoring that goes on or filtering, if you will}{J}{C1-T2}, and another said that given their past experience working in a formal ethics team, \pquote{we're all peers [...] I knew I could share freely in front of this group of people}{K}{C1-T2}.
Thus, while hierarchies may impact what some team members feel they can discuss, specific team norms may help to support all team members in speaking more freely, \blue{bringing forth perhaps the most direct example of how particular team dynamics affect how critique can be raised.}

\subsubsection{\blue{Awareness of teammates’ critical orientation}}\label{sec:critical-orientation}
Participants displayed awareness of who was in the room, and their ethical views and \blue{critical orientation. We define \textit{critical orientation} to mean an understanding of a teammate's (un)willingness to engage in ethical critique that may challenge existing project or business goals, as which may be risky~\cite{madaio2020co,widder2023power}.
We saw that participants' awareness of others teammates' critical orientation affected how they expressed themselves or whether they raised certain kinds of critique.}
One participant said this directly: \pquote{who is in the room can change the tenor of a conversation and can change the tenor of how you deliver critiques or hold back critiques}{E}{C1-T1}
She went on to say she'd frequently discuss concerns like ``privacy'' and ``fairness'' with all members of her group, but discuss concerns like AI displacing human labor with only a subset of them: \pquote{there's some people in the group, who are, whether by virtue of their discipline or their interests, are more attuned to [...] discussing things like labor}{E}{C1-T1}.

Others suggested that they habitually discuss AI ethics topics among their team, but that \pquote{the dynamics [...] probably would [...] be different if it was, like, any of us, with people from other teams}{RB}{C1-T1}, because they wouldn't have a \pquote{shared baseline}{RB}{C1-T1}.
Another participant stated \pquote{it would have taken me longer to sort of establish sort of as an internal feeling that people were sort of engaged in a discussion in good faith}{L}{C2-T1}.
Even those on AI ethics teams may not feel completely aligned with their direct team when shared views of the critical questions around AI are \pquote{not so sharply in focus}{D}{C2-T1}, as they were on their past teams.

Some were worried about particular consequences arising from raising critique around unfamiliar people, including that this could \pquote{impact [...] who I [can] collaborate with [...] some people can be really sensitive}{RM}{C1-T1}. One participant suggested that webs of social and collaboration networks are opaque in companies, leaving her unwilling to critique other researcher's projects.
%, because she doesn't know who knows who: \pquote{In industry, a lot of [this] is much more, like, invisible. I don't know who's worked together on stuff before. I don't know who's had whose backs at a meeting. I don't know who has supported each other in like in a review, or however people meet.}{E}{C1-T1}
Others reflected on raising specific topics during the game due to the (dis)comfort with their team. One company participant said she felt comfortable raising topics like worker exploitation because she knew \pquote{it's a group of [...] like-minded people}{AC}{AC}.

In a different example, an activist participant felt pressure to stay ``on topic'' and raise critical points, because her fellow players were such a \pquote{critical group of people. I might [otherwise] have been goofier in playing a card game. [...] more like trying to[...] just like fuck around}{A}{AC}.

% \subsubsection{(merge with above as counterpoint) Backlash}

% Others feared backlash if they were to express views different than what they saw as a public sentiment increasingly critical of AI, and its relationship to the automation of more kinds of work. 
% Others, who was a research scientist in machine learning, noted \pquote{I speak to these [artists,] right, the things that are that they create is something that DALL-E would create, you know, like instantly, right?}{RM}{C1-T1} was and other things he positioned as automating work, but explained that \pquote{If we put an artist into a conversation on DALL-E, [...] I would probably not voice my opinion [that the existence of these systems are] perfectly fine}{RM}{C1-T1}. 

% He also noted how he has become less comfortable talking about his work, for example at 
% a \pquote{restaurant, people overhearing us without context, I would probably stay more silent.}{RM}{C1-T1}, but how \pquote{five years back, like, 10 years back, I was, you know, like, I was bragging that this is what I this is what I work on}{RM}{C1-T1}
% Noted that it is plausible that his work may one day lead to automating work, and that he felt it was better his company do that rather than competitors, he reflected how he often didn't feel comfortable discussing ethical issues here in most contexts because \pquote{it's really a double-edged sword [...] you take one stand, or the other [...] you're sure to be slaughtered. }{RM}{C1-T1}
% This AI developer feared they may face some backlash when discussing AI ethics around people who may see his work. 

\subsubsection{Personal attributes and status hierarchies}\label{personal-atributes}
Participants recognized and discussed how gender, seniority, and level of technical experience affected the status one might have in a particular room, and thus the license with which they felt able to raise critique, or affected group dynamics in such a way that made raising critiques feel more or less possible.

We observed that age and gender affected perceptions of who is able to speak up. For example, one participant, lamented that his younger colleague \pquote{was not really talking up [speaking up]. He was not grabbing time}{RM}{C1-T1} because \pquote{he is really young. And [...] he joined very recently}{RM}{C1-T1}, In contrast, another very senior participant joked \pquote{you know me, I talk about anything. Maybe when I was younger, I might have been more cautious}{RB}{C1-T1}.
Another participant whose play session included only female-presenting people suggested \pquote{there's a different way these conversations happen in all-female groups than when there are other genders present [...] men take up space in particular ways}{R}{C2-T1}.
Such comments suggest how age, seniority, and gender may affect perceptions of who can or should speak up. 
%Another participant, the most senior in his play session, joked self-referentially:  

Participants also spoke about their roles within engineering organizations and their backgrounds.
One woman-identified participant with a non-engineering background stated: 
% \pquote{ I'm the new kid on the block. .. }{E}{C1-T1}, 
\pquote{if an engineer seems to be saying something that I think is wrong, I don't know, he's an engineer, and he's been here 20 years, maybe I'm wrong}{E}{C1-T1}, 
suggesting how seniority, ``engineering'' expertise, and perhaps gender, may impact who is perceived as ``wrong'' in company contexts. 
Participants in the activist group also questioned whether they should engage in critique while not having an engineering background.
A female-identified member of the activist group suggested her lack of computer science expertise may be a shortcoming, saying
\pquote{if I was in the room with people who were developing AI, I might feel uncomfortable just because I don't have the same depth of knowledge on the topic as they do}{}{AC}.
Another participant felt they might not qualify to participate in this study  \pquote{one of the requirements in [study criteria] was to be like in an engineering field. So I was like, I'm not that}{}{AC}.
\blue{This theme relates both familiar team dynamics, in the way that  particular identities (\ie gender, age) were looked on with credibility, but also organizational dynamics more salient to tech companies where those with engineering expertise are seen as most credible.}

 % She noted how ``jargon''  \pquote{would just be totally lost on me [...] or concepts that I never had never learned about. Because we all have different scopes of knowledge}{R-Act}
Some participants, many of whom had graduate educations, also reflected on the status of those with academic backgrounds and how this can quell critique in AI ethics discussions:  \pquote{whenever there's some very senior professor speaking [...] people don't speak out against them [...] people tend to agree}{RM}{C1-T1}.
One participant noted that their personality and the amount he speaks may lead others to agree too quickly, ``overpowering others'' \textit{(C1-T1)}.

\blue{
\subsubsection{Summary}
In this subsection we have shown the various ways that the particularities of who is in the room affects one's license to critique~\cite{christensen2017license} in discussions of AI ethics. This includes power dynamics resulting from one's manager being present, awareness of the critical orientation of colleagues in the room, and how personal attributes like age, gender, and expertise affect perceptions of who is able to raise critique.}

%saying 
%\pquote{you don't want to interrupt, you better be silent.}{RM}{C1-T1}. 
% Further, participants with a Ph.~D.~ A participant in a different group acknowledged she had a Ph.~D.~: \pquote{there's a difference in terms of how much schooling folks have had}{R}{C2-T1}, but that in her view, it is more ``egalitarian'' in practice. 

% \pquote{looking at the cards again, just making sure like if anything's came up, no, not really. I mean, I don't think any of these cards were controversial. They look like they were all, you know, like, pretty much similar. If one of us, were to be a person who was [the workers required to use our AI system], right? I wouldn't have brought something up. Right. But we are all we all work on the same project, developing the same things, looking at the same ethical problems and so on, right.}{RM}{C1-T1}

% \pquote{in saying something about, let's say, automating some of the things that a technician would do. Right. While, you know, like the, like, while the intention that I have is, I mean, like, you know, if this, going to happen, it's better Intel and our team, rather than, you know, like Microsoft doing these things}{RM}{C1-T1}

%\pquote{I used to browse academic Twitter [...] And in the discussions that were very, very heated in some cases. And it gave me the pressure to if nothing else thought that it would take time for people to start thinking about these things seriously. And I also came to my senses and stopped looking at Twitter.}{L}{C2-T1}

\section{RQ2: How do AI ethics discussions
unfold while playing a game oriented toward speculative critique?} 
\blue{

In this section, we examine how affordances from the game context appeared to affect how conversations unfolded. Following from our earlier discussion on how notions of scope limit license to critique (Sec.~\ref{sec:scope}), we firstly show how the introduction of randomness, discussion about harms beyond a particular system, and the creation of a hypothetical context allowed a wider scope of discussion.
\red{However, we also show} that participants felt that this expanded scope in a hypothetical context would not transfer back to discussions of actual projects, which are treated as compliance exercises and risk painting one's project in a negative light. 
Secondly, we demonstrate how the game enabled teammates to learn more about each other by creating a space to socialize, to be vulnerable, to learn about other's critical orientation, and to find ``allies'' for future discussions of ethics back in one's real work environment.    
}
\blue{Here, we rely primarily} on observations and recordings of the game session. 
We examine how the game was able to expand scope, how participants remixed rules, and how they used the game context as an opportunity to learn about teammates' past experiences and critical orientation.  

% in a team? alternatives: how does a game context impact AI ethics discussion in a team? what interactions does the game context afford?

\subsection{Expanding scope}\label{sec:expanding-scope}
\subsubsection{Randomness as scope expander}
Participants found that the randomness provided by the cards and game rules could be a valuable way to expand their conversations and critiques, especially beyond what they might normally discuss.
% Relatedly, participants suggested ways that affordances from the game enabled them to expand discussion beyond what one participant called a ``rut''. 
One participant whose work focuses on the ethical challenges of content recommendation reflected that  
\pquote{it was cool to [be] outside of the [content] recommendation space for a second [...]
Because you can get into a rut [and] having like a new [example] helps you see some of the gaps [...] in your own thinking [that you're ] habituated to}{R}{C2-T1}.
Participants from other sessions corroborated this, one stating \pquote{the format of giving responses [cards] ends up, like, forcing you to make connections that maybe you wouldn't have thought about before}{L in session}{C1-T2}. Another mentioned that a format where \pquote{there's not really a correct answer}{L}{C2-T1} is one she hadn't considered before, but noted that it \pquote{got a lot of us thinking in different directions}{L}{C2-T1}.

Participants also suggested that subjective interpretations of the same card served to expand the scope of discussion. One said this directly: \pquote{people came up with things I didn't expect, despite looking at the same card}{L}{C1-T2}.
A member of the team from the activist group suggested that ``randomization'' helped expand scope which was useful for a different reason, as \pquote{usually when I have conversations like this they're about a very specific real thing, right there, like either something's happened in public in the news, or [...] that someone's working on something [concerning]. They're not necessarily, like, speculative}{A}{AC}, thereby helping drive conversations about speculative possible futures that need not be reactive to any particular news event.
Other participants also expanded the scope of discussion by integrating parts of their own lived experiences during in-game discussions, integrating discussion from outside of the particular set of cards at hand.  
One participant referenced how technology had changed street culture in her native India by putting the ``juice man'' on the corner out of business as people moved to app-based delivery services, another relayed about protests against visual noise wrought by advertising on metro trains in her native Saint Petersburg, a third relayed how her native Berkeley was ``awash in Kiwibots'' with their ``pixel heart'' eyes, and fourth spoke about their experiences on a team where a robot had physically harmed someone. 
As one member of the activist group reflected: \pquote{all of us had such a different frame of reference}{F}{AC}, and people appeared to feel able to speak from this frame of reference throughout game play.

\subsubsection{Thinking beyond the product}
Participants also found that the game led them to consider scope beyond the product and towards second or third-order harms.
One participant stated \pquote{[we] were thinking like a couple of steps ahead [to] society at large, whereas [discussions] in practice tend to be about a more narrow, so like [...] how is this product harming users in ways that are measurable and quantifiable?
}{R}{C2-T1}
Similar sentiments were echoed by members of the activist group. 
For example, one participant reflected that in her life, she usually discusses concerns about AI within the context of a specific AI system ``actively happening'' in the present or recent past, and appreciated the opportunity to talk about future possibilities: \pquote{[our discussions] were more theoretical in the sense that we weren't talking so much about a specific form of AI [...]
% and how we've developed [it] over time [...]
%, feel often in my [life] outside of that game, when I talk about AI is in reference to something that is actively happening, whether it's something like Chat-GPT, or like the, like, the predictive risk modeling. 
So it was interesting to kind of talk about it in a more intangible way. Although it's always about, you know, kind of predicting the future in an intangible way}{R-Act}{AC}.
Another member of the activist group echoed this separately in their own follow up interview, suggesting that their in-game discussion spoke to \pquote{these cultural intangibles that really got to, I think, the deeper root of some of our concerns}{F}{AC}. 

% This relates to what Christiansen \etal refers to as discursive ``closure by the past'', where consideration and discussion of critique is bound by past discussions or concerns, making it harder to think about possible futures. 

\subsubsection{Hypothetical situations as an ``innocuous'' context for discussion}\label{sec:innocuous-context}
Several participants brought up how the hypothetical context of the game provided an ``innocuous'' context~\cite{scott1990domination} to raise critique that they felt may have otherwise been too socially costly to raise. 
Given randomly drawn prompts and dealt response cards provided, one participant suggested if the ``structure of the game'' ``pushes'' one to \pquote{bring up things that you [otherwise] wouldn't feel comfortable bringing up}{L}{C1-T2} then \pquote{in that context, it probably does make it easier}{L}{C1-T2}.
Reflecting on other hypothetical interventions she's participated in before, this participant also reflected that this makes it easier for people not just to raise critique they may have but be nervous to raise, but accept and themselves raise critique of things \textit{similar} to their own work while being less ``defensive'':
\pquote{once we did it in a hypothetical sense, people were looking at this and going `oh, okay, well, yeah, it's not about whether we intended for something to go wrong [...] things can really go wrong!'}{L}{C1-T2}, saying that this allowed people to ``disconnect'' from the frame of \pquote{`something that you're doing is incorrect', or `there's something unethical about your work'}{L}{C1-T2}, that when conversation is  
\pquote{taken away from the project that we were doing [...] everyone was very free}{RM}{C1-T1}.

Others corroborated this, suggesting that raising concerns \pquote{hypothetically in a game [...] is really nice because it's just a lower barrier [...] versus talking about a specific project which [...] is going to be much more serious and have potential real-life implications right as you bring up different concerns}{Session}{C1-T2}.
One said the point is to ``make assumptions'': \pquote{I saw this one [card] is, you know [...] I'm making a lot of assumptions here. But that I think that's maybe the point of some of these discussions}{Session}{C1-T2}.
Another participant remarked how she appreciated the opportunity to \pquote{take ourselves out of, you know, okay, `this is a real product that we have to provide actionable guidance and feedback on' to [instead] `okay, let's just have our, you know, brain flowing to think about all the possible what ifs, what could go wrong with this scenario'}{K}{C1-T2}.

\subsubsection{Hard to transfer from hypothetical context to real world action}\label{sec:transfer}
However, we found that this hypothetical context may make it difficult for in game discussions to transfer to real world action. 
Importantly, this raises questions about whether discussions rooted in in-game hypotheticals can spur real-world action. 
Reflecting on past AI Ethics trainings based on hypotheticals, one participant found them effective, but noted: \pquote{The minute you start talking about their projects, you see a very different behavior. [...] They're very concerned about these projects showing up in a negative light. And [...] people start to become more defensive. They don't expand into all the things that can go wrong}{L}{C1-T2}.

Revisiting a quote from the first part of this section, one participant said: 
\pquote{hypothetically in a game [...] is really nice because it's just a lower barrier [...] versus talking about a specific project which just by nature is going to be much more serious and have potential real life implications right as you bring up different concerns}{Session}{C1-T2}. Among these ``real-life implications'' may be the idea of real world action, such as through existing compliance processes. However, one recommended against this: \pquote{I could see possible resistance is if it's seen as a checklist activity. So if it's perhaps tied into like a compliance process, and like, you must do this before your product goes out the door, then there could be some resistance there}{Session}{C1-T2}.
This suggests that it may be difficult to fuse the hypothetical context created with the game with an integration with requirements for mandated changes to actual products. 

A participant in a different group echoed this, reflecting on the good conversation from her groups' play session, and wishing there would be a way to translate this into action: \pquote{My complaint [...] with team based [...] conversations... like when two people talk  [...] the whole is greater than the sum of the parts. [...] But translating that thing that is made into something that is captured and can be operationalized [...] has been a consistent issue}{E}{C1-T1}. This was also apparent in the words some participants used to refer to the session, one calling it a \pquote{non-work space [...] almost like a team building exercise}{R}{C2-T1}, which appeared to set the expectation that this is not the context from which immediate or actionable changes to the product or process in work contexts would arise.

 \subsection{Learning about teammates}\label{sec:learning-teammates}
\subsubsection{Vulnerability and space to socialize}\label{sec:vulnerability}
Participants spoke about how the game context made certain conversations possible that they felt otherwise unable to have. 
In one instance the Response card ``Random Crashes'' prompted a participant to share that he had previously worked on a robot which had killed its operator. 
This was the first time he had shared this with his teammates despite them working on physical systems and frequently discussing safety and ethics concerns. 
In follow-up interviews, his colleagues reflected on this: \pquote{[he] shared with us that he was working in this factory, where actually a robot did a ``random crash'' and [...] killed somebody. [...] It was impact[ful], like it was, it was really shocking. Like `Oh, wow' like he was part of it. Like he was there}{G}{C1-T1}.
Another participant suggested that with an \textit{``all-audio [meeting] culture''} that it was%
%}{E}{C1-T1} 
\pquote{nice to get that kind of space where we could more like, really talk about things we were seeing or things that we were thinking about [that were] not necessarily constrained by our work}{E}{C1-T1}.
\blue{This speaks how perhaps team dynamics within companies, by default, do not enable vulnerability nor space to talk beyond work tasks, both of which appeared to be helpful when discussing ethics.}

\subsubsection{Learning others' critical orientation and finding allies}
Apart from sharing sensitive past experiences, some spoke about how the vulnerability prompted by the game created a unique opportunity to learn about a teammate's critical orientation.
By \textit{critical orientation}, we refer both to their values and perspective on ethical issues in technology, but also their willingness to critique project's goals or company incentives, when this may be in conflict with the former. 

For example, one reflected how a \pquote{non-work space, but [where we were] still be able to have conversations that are adjacent to what we're doing [...] helped me see that we're more or less all on the same page [and] who my allies are in this fight}{R}{C2-T1}.
On a similar point, one participant in another company
stated how she had previously discussed more critically oriented topics, such as labor displacement, with only some of her coworkers, but had\pquote{probably self-selected out of discussing certain things with [other] folks due to [their] backgrounds [or] presuming that they're not interested}{E}{C1-T1}.
However, reflecting on the game session, she relayed how she appreciated hearing from teammates \pquote{with whom conversations can be very tight and narrow, to hear them pontificating a little more, engaging in [an] imaginative exercise. [...] I've only ever heard them talk about dialogue prompts [so] it can be easy to assume [that they] don't think the same way that I do [... and because of this,] theory of mind can be difficult to achieve}{E}{C1-T1}.
In her view, this game provided an opportunity for her to learn how more members of her team felt about more critically-oriented topics. 

In a follow-up interview, one of the members of the activist group discovered that she had similar concerns to a member of the group she had only met briefly, and after playing the game noted: 
\pquote{I really like their perspective [,...]some things that they said [...] made me want to talk to them further}{A}{AC}.

We note that exposing one's critical orientation, especially around those one does not know well, may be a risky endeavor: one risks being labeled a troublemaker~\cite{ahmed2021complaint} or facing career repercussions~\cite{madaio2020co}. \blue{Team dynamics which better enable one to feel safe exposing views that may challenge business logics may alleviate the sense of isolation that past work shows people can feel when holding such views~\cite{widder2023power}, or the ``malaise'' that may ``permeate'' through organizations when this is more widespread~\cite{su2021critical}}
We return to this further in Section~\ref{sec:not-direct-critically-alligned}, below. 

\section{Discussion}
% \subsection{Power, and its effect on how one exposes their critical orientation}
Our results show that in the game and in ordinary work, discussants seek to understand and display sensitivity to both the differentiated power and critical orientation of their discussion partners, which they use to modulate AI ethics issues they choose to raise and how they present them. 
When one's boss or colleagues of unknown critical orientation are in the room, people may be less willing to raise critique. 
This echoes the work of James Scott, demonstrating how people employ ``public transcripts'' when those with power over them are present, but use more frank offstage talk when speaking to teammates they trust~\cite{scott1990domination}. 
Additionally, a variety of factors affect people's perception of their own status, such as their seniority in the team or their proximity to engineering knowledge, in turn also affecting their willingness to raise critique. 

The most straightforward implication of this finding is that those designing future AI ethics interventions intended to be used in a group discussion context must attend to the differentiated power relations of discussants. 
Explicit attention to this may include exercises for a group discussion to begin by reflecting on these, reflexively discussing these as a group, or even simply an enumeration of what kind of power relationships (\ie, boss/subordinate, as well as those related to age, seniority and gender) to look out for. 
Naming these things will not level them, but doing so is already more attentive to power dynamics than many existing AI ethics interventions. 
Our work joins a great deal of other work~\cite{birhane2022forgotten,gansky2022counterfacctual,lee2020power,widder2023power,johnson2019ai} which makes clear that research on AI Ethics must be more attentive to the differentiated power relationships between those that may use, request, or engage in proposed AI ethics interventions or discuss AI ethics issues, \blue{and how these power relationships are ingrained in organizational conventions and culture.}
If future empirical work examining how those discussing AI ethics in any context does not attend to power in their analysis, such work risks missing major determinants of any apparent agreements or disagreements that may arise. 

\subsection{A hypothetical game context \blue{is unlikely to lead to direct change}, but may help find critically-aligned allies}\label{sec:not-direct-critically-alligned}
\blue{We join past work in discussing how tech industry logics such as technological solutionism affect ethics initiatives ~\cite{metcalf2019owning}, and other topics intersecting with ethics. These include privacy, where some worry that ``advocating for privacy might be indirectly misaligned with career incentives''~\cite{leedon}; accessibility, where this is not seen as a first-order concern and only attended to if customers ask for it~\cite{widder2023power}, or studies which show that implementing accessibility is an individual rather than an institutional responsibility~\cite{bi2022accessibility}; and sustainability, where the ``endless pursuit of achieving higher model quality has led to the exponential scaling of AI with significant energy and environmental footprint implications''~\cite{wu2022sustainable}, especially in the context of increasingly Large Language Models~\cite{bender2021dangers, gururaja2023build}.}

Our work casts doubt on whether an innocuous context, such as those created by games, may enable discussions that \blue{successfully challenge these logics and} transfer to changes in a team's real-world context. \blue{In short, \red{our findings suggest that} games are unlikely to present opportunities to escape the power dynamics that shape AI ethics discussion in ways that lead directly to project changes.} 
While Scott's suggestion that ``rumors, gossip, folktales, songs, gestures, jokes'' are the places where people may demonstrate dissent more freely by ``hiding behind anonymity or behind innocuous understandings'' ~\cite{scott1990domination}, \blue{raising questions about whether} game-based AI Ethics interventions may expand the scope of what is sayable ``on-stage'' and create such contexts to make dissent safer, our results tell a more complicated story. 
As we detail in Section~\ref{sec:innocuous-context}, participants spoke about how they felt able to speak freely specifically because the context was hypothetical: not connected to a particular project and not tied to a particular ``compliance process,'' as one participant said, which may demand politically difficult or time-consuming changes to one's product. 
This gap between the hypothetical context and ``real-life implications,'' as one participant put it,
% (as one participant put it, implications which they sought to avoid) 
is both a powerful attribute of the intervention---it is the feature that made specific conversations possible that were not before---\textbf{but also a powerfully limiting factor of the intervention, in that this gap was seen as being maintained by \textit{not} implying any change outside of the hypothetical context. }

Our findings therefore cast doubt on \blue{the utility of games and other hypothetical interventions to create space for} discussions or agreements that transfer back to action in business contexts, where this would imply real work, real shifts in direction, or sign-off from higher-ups. 
\blue{Given this, our findings suggest that interventions such as games \red{may be} unlikely to meaningfully intervene in power dynamics in ways that directly spur real-world action, \red{given organizational logics and policies}.}
For example, when proposing the game we study, Martelaro \etal claim that a ``little lightheartedness can promote more productive conversation about otherwise negative topics''~\cite{martelaro2020could}, 
and Ballard \etal found that ``having a serious conversation about ethics and technology in the context of a game creates space for difficult or uncomfortable conversations. 
Within this conversation, the use of design fiction to create discursive space [...] deflects blame or charges of irresponsibility in actual settings with actual harms''~\cite{ballard2019judgment}.
While our results suggest that discursive space may indeed have been created, it is still unclear and unknown how such conversation may lead to averting actual harm from real work.
% our results suggest that discussion within this space may not lead directly to averting ``actual harms'' in actual contexts.

This being said, our results suggest a more subtle and perhaps enduring mechanism of action for games to shift organizational realities: finding allies by developing an understanding of their critical orientation through gameplay.
Our results suggest that the hypothetical context fosters vulnerability, such as through sharing sensitive past experiences working on AI systems that had caused deadly harm.
Such stories help reveal parts of team members' critical orientation and allow others to learn about their critical orientation (see Sec.~\ref{sec:critical-orientation}).
When these personal understandings and relationships transfer to real-world contexts, this may help form coalitions to address real-world ``actual harms.'' 
Scott emphasizes continuity between the two ``stages''~\cite{scott1990domination} he proposes, and relationships that form ``off-stage'' appear to be the conduit towards enabling ``on-stage'' solidarity. 
% Our results showed how the hypothetical context created by the game enabled participants to feel comfortable being vulnerable, for example .
Another participant discussed how they had felt comfortable discussing possible labor displacement implications of AI systems they were themselves building, conversations which they had not previously had with certain members of their group, presuming some were not interested in such topics. 
Reflecting that in ordinary work contexts, \textit{``theory of mind can be difficult to achieve,''} she relayed how she appreciated learning more about her teammates on topics they didn't usually discuss through this ``imaginative exercise''. 
Another participant in a different group reflected on how this game helped her learn \textit{``who my allies are in this fight.''}
While strengthened social ties, or one or two more allies may seem small, ``if subordinates are entirely atomized, of course, there is no lens through which a critical, collective account'' can emerge~\cite{scott1990domination}, and we join Scott to suggest that collective accounts are where solidarity begins. \blue{It is nonetheless important to note that games are, at best, a modest intervention, perhaps providing a possible context for \red{solidarity} to arise, among scant alternative spaces.}

%pg number for above is 134

% So, rather than creating space for discursive openness, and hoping that discussion from that transfers to real-world action, our results show how the innocuous context created by the game gives people an opportunity to learn about the critical orientation of their teammates -- and feel more comfortable revealing their own. 
While formal AI ethics activities such as checklists may be able to ``empower \textit{individual} advocates'' (emphasis added) by legitimizing a particular issue~\cite{madaio2020co} contained on a narrowly scoped checklist, intersubjectivity developed through gameplay may enable individuals to better know one another---who their \textit{``allies are'' }in the words of one participant---from which a broader collective to raise critique may be fashioned, less constrained by the discursive limits of any particular standard~\cite{christensen2017license}. 
Past studies demonstrate the severely limited ability of employees to raise concerns beyond a very narrow scope, and instead suggest that future ``ethics interventions, research, and education must expand from helping practitioners merely identify issues to instead helping them build their (collective) power to resolve them''~\cite{widder2023power}, and our results suggest that ``innocuous'' contexts (\ie~\cite{scott1990domination}) created by games may provide space for collective power to begin to form. 
In organizational psychology, this concept is termed  ``cross-understanding'', defined as ``the extent to which team members understand the other members’ mental models''~\cite{janardhanan2020getting}. While the literature on this construct often focuses on the impact of cross-understanding for narrower questions of ``product quality''  and avoids questions that members would find ''technically, politically, or otherwise unacceptable''~\cite{huber2010crossunderstanding}, parallels may be drawn beyond quality and features, to questions of product ethics.

Feminist theory helps illuminate the distinction between hypothetical or ``innocuous''~\cite{scott1990domination} contexts enabling real-world changes directly versus enabling stronger ties, which \blue{may then become a basis from which action may then arise.} 
Donna Haraway argued that we ought not to suppose that there is a ``view from above, from nowhere'', and thus that trying to suppose a context that can create one, is both unlikely to succeed and may be harmful~\cite{haraway1991situated}. 
Extending Haraway's argument, Lucy Suchman argues that responsibly developing technology must be a ``boundary-crossing activity, taking place through the deliberate creation of situations that allow for the meeting of different partial knowledges''~\cite{suchman2002located}. 
Rather than theorizing games as separate safe spaces from which to speak from nowhere, we suggest that games may be \blue{a modest but deliberately created opportunity for} different partial knowledges to meet, learn what they have in common, and enable ``collective knowledge of the specific locations of our respective visions''~\cite{suchman2002located}, from which durable coalitions and collectivities for action may arise. 
Drawing on both Haraway and Suchman, Widder and Nafus show how social ties, responsibilities, and concerns---developed outside of engineers' assigned duties---are the basis for the (little) AI ethics work that does get done, and innocuous contexts created by games may enable non-work contexts for these ties to form and strengthen~\cite{widder2023dislocated}. 
% In some sense, while a participant cast this as a ``non-work space'' and ``a team building exercise'' originally appeared to diminish the power of these spaces to enable change, it is perhaps precisely ties developed in these spaces that enduring coalitions may start to form, who may advocate for ethically driven changes within ``work'' or ``product'' spaces. 

This has implications for game design research, especially if intending to intervene in group dynamics for prosocial ends, particularly in contexts like workplaces with built-in power hierarchies. Such work may consider framing their game as an opportunity for relationship and coalition building more so than a context where direct changes to real practice will arise. This may include examining the effect of any such intervention and examining contingencies on the durability and outcomes from any resulting relationships formed, over a longer time span.

\subsection{``Out of scope'' as a rhetorical device to softly dismiss critique}
% Wong and Madaio analyze \cite{wong2022seeing}

Our results 
% illustrate how those warnings of discursive closure in responsible AI, as theorized in past work on a larger scale, play out in practice in the teams we study. 
show how notions of ``scope''---what a team believes it can or will take action on---are constructed and maintained, how this limits what is considered acceptable in AI ethics conversations, and ways that participants sought to say things outside of these bounds (see Sections~\ref{sec:scope} and \ref{sec:expanding-scope}). 
Firstly, we show how \textbf{individuals compartmentalize ethics} in ways that limit what they perceive as in scope during AI ethics discussions, with participants from companies suggesting that out-of-work experiences or passions are not in scope for AI ethics discussions.
In contrast, those in the community activist group did not feel this way. 
Additionally, some of our participants discussed how labor is divided in ways that leave ethics to be the primary remit of one team member, leaving others feeling that ethics issues are beyond their own ``scope.'' 
Some of our participants also segment critique between projects, 
% not referencing critique between the different projects they work on 
in order to avoid perceived career consequences when working with different team members. 
In these ways, the wholeness of any individual's perspective is itself compartmentalized, leading to a narrowed scope of discussion when teams meet. 
This elaborates what Widder and Nafus argue~\cite{widder2023dislocated}, but demonstrates how ethics is modularized between team members and within individuals as they choose to bring 
% while everyone has a ``partial perspective'' in that no person has access to a ``view from nowhere'' god-like perspective, this shows how people also bring 
only \textit{fragments} of their own partial perspective to these discussions. 
Secondly, our results illustrate how notions of \textbf{efficiency become a scope limiter}, casting a subjective assessment of priorities in the more objective language of ``scope''.  Participants' direct references to their calendar and scarcity of time, and to less direct notions of relevance or framing some harms as ``unlikely'', lead to a situation where only possible harms perceived as most relevant, or most likely, are seen as most in scope and thus most legitimate to raise for discussion. 
% This is in contrast to the activist group we study who report this being more of 

Thirdly, given these time pressures, teams report how discussion is most often scoped towards \textbf{that which feels actionable, often technical changes,} in line with how technosolutionism operates to limit discussion to immediately actionable fixes. 
This is evident in how participants describe their workflow: being presented with particular systems they are supposed to evaluate for ethical issues, and propose ``mitigations'' to solve said issues, and this frame makes it harder to discuss how the systems they're evaluating relate to one another, or how they may require ``systemic changes'' that this frame does not present as part of their agency to discuss.

Finally, our results suggest how \textbf{scope can be tested or expanded}, in how affordances from game-like interventions such as randomness may give social permission to do this (see Section~\ref{sec:expanding-scope}), and how people may employ rhetorical moves to frame ethics questions as larger scoped product questions. 

We theorize scope as a softer way to dismiss critique. 
This functioned as an instance of problem closure~\cite{humphreys2005reframing}, whereby ``rhetorical process through which relevant social groups perceive their problems with an artifact to be solved or closed'', but in a way that softly dismisses those who may wish to keep it open or believe it to be unsolved. 
\blue{Our data shows that} casting an issue as ``out of scope''  merely says that it is beyond the team's remit or practical ability to act on, without forcing one to contend directly with the issue raised. By avoiding the need to deny the validity of an issue outright but instead suggesting it is out of scope, one can avoid dismissing the validity of a colleague's sincerely raised ethical concern.
This is a particular instantiation of \textit{jurisdictional stasis} -- a concept in rhetorical analysis with its roots in classical Greek, but with more modern adaptations to questions of ``moral decision making or ... practical concerns''~\cite{tomlinson2020stasis}.  
Instead of arguing that an issue is false, arguments based on jurisdictional stasis question the ``jurisdictional appropriateness of the issue''~\cite{tomlinson2020stasis}, that is, whether an issue is within the jurisdiction or scope of a particular group or team. 

While not always described as such, many scholars have theorized how standards, principles, and toolkits seeking to guide organizational behavior toward ``pro-social'' are discursively closed. 
In their analysis of environmental sustainability standards, Chistensen \etal ~\cite{christensen2017license} demonstrate how they risk discursive closure: \textit{closure by the past}, where responses to future problems are limited by standards developed for past concerns; \textit{closure by design}, where overly-prescriptive standards leave no freedom for adaption and become a putative ``seal of approval''; and \textit{closure by routinization}, where standards are solidified into organizational processes in ways that are difficult to change. 
In an AI ethics context specifically, Greene \etal analyze AI ethics statements of principles, examining how they ``legitimate (and delegitimize) certain practices'', finding in part that by focusing on how to design AI systems rather than the business practices they enable, they frame ``business practices [as] being discursively `off the table''', implying that ```better building' is the only ethical path forward'''~\cite{greene2019better}. 
Keyes \etal satirically demonstrate how narrowly scoped ``Fair, Accountable, Transparent'' design principles scope scrutiny to system design, warning against ``treatment of ethics as a series of heuristic checkboxes that can be resolved technically'' and thereby avoiding engagement with ``wider societal issues''~\cite{keyes2019mulching}.  

% Link to``Scope creep'', in software engineering.

This suggests that designers of future AI ethics interventions ought to see the risk of discursive closure and deploy particular ways to reduce this risk. 
Our results suggest particular design affordances that may help do this.  
While the designers of the 2020 Microsoft AI Fairness checklist recognized the risk of discursive closure in that they included disclaimers in the checklist's extensive preamble like ``Undertaking the items in this checklist will not guarantee fairness. The items are intended to prompt discussion and reflection''~\cite{madaio2020microsoft}, our results suggest that more than written disclaimers or warnings are needed to avoid discursive closure.
More extensive changes to an intervention's form and including deliberate design affordances, such as randomization, are needed to resist this closure. 

\section{Conclusion}
Past work has sought to design AI ethics interventions--such as checklists~\cite{madaio2020co} or toolkits~\cite{bird2020fairlearn}--to help practitioners design more ethical AI systems. 
However, other work demonstrates how these interventions~\cite{wong2022seeing} and the principles they're based on~\cite{greene2019better} may serve to instead limit critique to those addressed within the intervention, while rendering broader concerns illegitimate, \blue{and how core logics of the tech industry make raising ethical concerns that reach beyond technological solutionism or market fundamentalism~\cite{metcalf2019owning} to challenge business practice~\cite{widder2023power} extraordinarily difficult.}

In this paper, we examined how teams discuss AI ethics issues by drawing on past work examining how standards in other contexts enact discursive closure~\cite{christensen2017license}, and on how power relations affect whether and how critique is raised~\cite{scott1990domination}. We recruit three corporate teams, and one activist team, each with prior context with one another, to play a game designed to trigger broad discussion around AI ethics, and firstly use this as a point of contrast to trigger reflection on their teams' past discussions, examining \blue{organizational and team dynamics} which may affect their ``license to critique'' in AI ethics discussion. We then report on how particular affordances of this game may influence discussion, paying particular attention to hypothetical games as a viable mechanism for real-world change. 
We discuss how power dynamics in a group and notions of ``scope'' affect whether people may be willing to raise critique in AI ethics discussions. Our finding suggest that games may not be able to lead to direct change, but may be more likely to allow members to find critically-aligned allies for future action. 

\section{Acknowledgments}
We are grateful to Wendy Ju, who helped develop the \textit{What Could Go Wrong?} card game with Nikolas Martelaro.
We thank Justin Deyo, Katie Ciez and Grace Klein at the Carnegie Mellon University Block Center for Technology and Policy for for hosting early pilot tests, 
and to student research assistants Sherry Chen, Eunbea Ji and Wan Tong Li for helping run these tests.
We are grateful to our anonymous participants, Michael Madaio, Dawn Nafus, Lucy Suchman, and Julie Martinson Widder, who gave feedback on earlier drafts, as well as the support of the Digital Life Initiative at Cornell Tech.

\bibliographystyle{ACM-Reference-Format}
\bibliography{sample-base}


\begin{thebibliography}{66}

%%% the Web address.

\ifx \showCODEN    \undefined \def \showCODEN     #1{\unskip}     \fi
\ifx \showDOI      \undefined \def \showDOI       #1{#1}\fi
\ifx \showISBNx    \undefined \def \showISBNx     #1{\unskip}     \fi
\ifx \showISBNxiii \undefined \def \showISBNxiii  #1{\unskip}     \fi
\ifx \showISSN     \undefined \def \showISSN      #1{\unskip}     \fi
\ifx \showLCCN     \undefined \def \showLCCN      #1{\unskip}     \fi
\ifx \shownote     \undefined \def \shownote      #1{#1}          \fi
\ifx \showarticletitle \undefined \def \showarticletitle #1{#1}   \fi
\ifx \showURL      \undefined \def \showURL       {\relax}        \fi
% invisible to TeX
\providecommand\bibfield[2]{#2}
\providecommand\bibinfo[2]{#2}
\providecommand\natexlab[1]{#1}
\providecommand\showeprint[2][]{arXiv:#2}

\bibitem[Ahmed(2021)]%
        {ahmed2021complaint}
\bibfield{author}{\bibinfo{person}{Sara Ahmed}.}
  \bibinfo{year}{2021}\natexlab{}.
\newblock \bibinfo{booktitle}{\emph{Complaint!}}
\newblock \bibinfo{publisher}{Duke University Press}.
\newblock


\bibitem[Ahmed and Swan(2006)]%
        {ahmed2006doing}
\bibfield{author}{\bibinfo{person}{Sara Ahmed} {and} \bibinfo{person}{Elaine
  Swan}.} \bibinfo{year}{2006}\natexlab{}.
\newblock \showarticletitle{Doing {{Diversity}}}.
\newblock \bibinfo{journal}{\emph{Policy Futures in Education}}
  \bibinfo{volume}{4}, \bibinfo{number}{2} (\bibinfo{date}{June}
  \bibinfo{year}{2006}), \bibinfo{pages}{96--100}.
\newblock
\showISSN{1478-2103}
\urldef\tempurl%
\url{https://doi.org/10.2304/pfie.2006.4.2.96}
\showDOI{\tempurl}


\bibitem[Ballard et~al\mbox{.}(2019)]%
        {ballard2019judgment}
\bibfield{author}{\bibinfo{person}{Stephanie Ballard},
  \bibinfo{person}{Karen~M. Chappell}, {and} \bibinfo{person}{Kristen
  Kennedy}.} \bibinfo{year}{2019}\natexlab{}.
\newblock \showarticletitle{Judgment {Call} the {Game}: {Using} {Value}
  {Sensitive} {Design} and {Design} {Fiction} to {Surface} {Ethical} {Concerns}
  {Related} to {Technology}}. In \bibinfo{booktitle}{\emph{Proceedings of the
  2019 on {Designing} {Interactive} {Systems} {Conference}}}.
  \bibinfo{publisher}{ACM}, \bibinfo{address}{San Diego CA USA},
  \bibinfo{pages}{421--433}.
\newblock
\showISBNx{978-1-4503-5850-7}
\urldef\tempurl%
\url{https://doi.org/10.1145/3322276.3323697}
\showDOI{\tempurl}


\bibitem[Bender et~al\mbox{.}(2021)]%
        {bender2021dangers}
\bibfield{author}{\bibinfo{person}{Emily~M Bender}, \bibinfo{person}{Timnit
  Gebru}, \bibinfo{person}{Angelina McMillan-Major}, {and}
  \bibinfo{person}{Shmargaret Shmitchell}.} \bibinfo{year}{2021}\natexlab{}.
\newblock \showarticletitle{On the Dangers of Stochastic Parrots: Can Language
  Models Be Too Big?}. In \bibinfo{booktitle}{\emph{2021 ACM Conference on
  Fairness, Accountability, and Transparency}}. \bibinfo{pages}{610--623}.
\newblock


\bibitem[Benders and Van~Hootegem(1999)]%
        {benders1999teams}
\bibfield{author}{\bibinfo{person}{Jos Benders} {and} \bibinfo{person}{Geert
  Van~Hootegem}.} \bibinfo{year}{1999}\natexlab{}.
\newblock \showarticletitle{Teams and Their {{Context}}: {{Moving}} the {{Team
  Discussion Beyond Existing Dichotomies}}}.
\newblock \bibinfo{journal}{\emph{Journal of Management Studies}}
  \bibinfo{volume}{36}, \bibinfo{number}{5} (\bibinfo{year}{1999}),
  \bibinfo{pages}{609--628}.
\newblock
\showISSN{1467-6486}
\urldef\tempurl%
\url{https://doi.org/10.1111/1467-6486.00151}
\showDOI{\tempurl}


\bibitem[Bi et~al\mbox{.}(2022)]%
        {bi2022accessibility}
\bibfield{author}{\bibinfo{person}{Tingting Bi}, \bibinfo{person}{Xin Xia},
  \bibinfo{person}{David Lo}, \bibinfo{person}{John Grundy},
  \bibinfo{person}{Thomas Zimmermann}, {and} \bibinfo{person}{Denae Ford}.}
  \bibinfo{year}{2022}\natexlab{}.
\newblock \showarticletitle{Accessibility in {{Software Practice}}: {{A
  Practitioner}}'s {{Perspective}}}.
\newblock \bibinfo{journal}{\emph{ACM Transactions on Software Engineering and
  Methodology}} \bibinfo{volume}{31}, \bibinfo{number}{4} (\bibinfo{date}{Oct.}
  \bibinfo{year}{2022}), \bibinfo{pages}{1--26}.
\newblock
\showISSN{1049-331X, 1557-7392}
\urldef\tempurl%
\url{https://doi.org/10.1145/3503508}
\showDOI{\tempurl}


\bibitem[Bird et~al\mbox{.}(2020)]%
        {bird2020fairlearn}
\bibfield{author}{\bibinfo{person}{Sarah Bird}, \bibinfo{person}{Miro
  Dud{\'\i}k}, \bibinfo{person}{Richard Edgar}, \bibinfo{person}{Brandon Horn},
  \bibinfo{person}{Roman Lutz}, \bibinfo{person}{Vanessa Milan},
  \bibinfo{person}{Mehrnoosh Sameki}, \bibinfo{person}{Hanna Wallach}, {and}
  \bibinfo{person}{Kathleen Walker}.} \bibinfo{year}{2020}\natexlab{}.
\newblock \showarticletitle{Fairlearn: A toolkit for assessing and improving
  fairness in AI}.
\newblock \bibinfo{journal}{\emph{Microsoft, Tech. Rep. MSR-TR-2020-32}}
  (\bibinfo{year}{2020}).
\newblock


\bibitem[Birhane et~al\mbox{.}(2022)]%
        {birhane2022forgotten}
\bibfield{author}{\bibinfo{person}{Abeba Birhane}, \bibinfo{person}{Elayne
  Ruane}, \bibinfo{person}{Thomas Laurent}, \bibinfo{person}{Matthew S.~Brown},
  \bibinfo{person}{Johnathan Flowers}, \bibinfo{person}{Anthony Ventresque},
  {and} \bibinfo{person}{Christopher L.~Dancy}.}
  \bibinfo{year}{2022}\natexlab{}.
\newblock \showarticletitle{The {{Forgotten Margins}} of {{AI Ethics}}}. In
  \bibinfo{booktitle}{\emph{2022 {{ACM Conference}} on {{Fairness}},
  {{Accountability}}, and {{Transparency}}}}. \bibinfo{publisher}{{ACM}},
  \bibinfo{address}{{Seoul Republic of Korea}}, \bibinfo{pages}{948--958}.
\newblock
\showISBNx{978-1-4503-9352-2}
\urldef\tempurl%
\url{https://doi.org/10.1145/3531146.3533157}
\showDOI{\tempurl}


\bibitem[Boyd(2021)]%
        {boyd2021datasheets}
\bibfield{author}{\bibinfo{person}{Karen~L. Boyd}.}
  \bibinfo{year}{2021}\natexlab{}.
\newblock \showarticletitle{Datasheets for Datasets Help ML Engineers Notice
  and Understand Ethical Issues in Training Data}.
\newblock \bibinfo{journal}{\emph{Proc. ACM Hum.-Comput. Interact.}}
  \bibinfo{volume}{5}, \bibinfo{number}{CSCW2}, Article
  \bibinfo{articleno}{438} (\bibinfo{date}{oct} \bibinfo{year}{2021}),
  \bibinfo{numpages}{27}~pages.
\newblock
\urldef\tempurl%
\url{https://doi.org/10.1145/3479582}
\showDOI{\tempurl}


\bibitem[Brucks and Levav(2022)]%
        {brucks2022virtual}
\bibfield{author}{\bibinfo{person}{Melanie~S. Brucks} {and}
  \bibinfo{person}{Jonathan Levav}.} \bibinfo{year}{2022}\natexlab{}.
\newblock \showarticletitle{Virtual Communication Curbs Creative Idea
  Generation}.
\newblock \bibinfo{journal}{\emph{Nature}} \bibinfo{volume}{605},
  \bibinfo{number}{7908} (\bibinfo{date}{May} \bibinfo{year}{2022}),
  \bibinfo{pages}{108--112}.
\newblock
\showISSN{1476-4687}
\urldef\tempurl%
\url{https://doi.org/10.1038/s41586-022-04643-y}
\showDOI{\tempurl}


\bibitem[Christensen et~al\mbox{.}(2017)]%
        {christensen2017license}
\bibfield{author}{\bibinfo{person}{Lars~Th{\o}ger Christensen},
  \bibinfo{person}{Mette Morsing}, {and} \bibinfo{person}{Ole Thyssen}.}
  \bibinfo{year}{2017}\natexlab{}.
\newblock \showarticletitle{License to {{Critique}}: {{A Communication
  Perspective}} on {{Sustainability Standards}}}.
\newblock \bibinfo{journal}{\emph{Business Ethics Quarterly}}
  \bibinfo{volume}{27}, \bibinfo{number}{2} (\bibinfo{date}{April}
  \bibinfo{year}{2017}), \bibinfo{pages}{239--262}.
\newblock
\showISSN{1052-150X, 2153-3326}
\urldef\tempurl%
\url{https://doi.org/10.1017/beq.2016.66}
\showDOI{\tempurl}


\bibitem[Clarke and Braun(2021)]%
        {clarke2021thematic}
\bibfield{author}{\bibinfo{person}{Victoria Clarke} {and}
  \bibinfo{person}{Virginia Braun}.} \bibinfo{year}{2021}\natexlab{}.
\newblock \showarticletitle{Thematic analysis: a practical guide}.
\newblock \bibinfo{journal}{\emph{Thematic Analysis}} (\bibinfo{year}{2021}),
  \bibinfo{pages}{1--100}.
\newblock


\bibitem[Conger and Wakabayashi(2019)]%
        {nyt2019retaliation}
\bibfield{author}{\bibinfo{person}{Kate Conger} {and} \bibinfo{person}{Daisuke
  Wakabayashi}.} \bibinfo{year}{2019}\natexlab{}.
\newblock \showarticletitle{Google Employees Say They Faced Retaliation After
  Organizing Walkout}.
\newblock
  \bibinfo{howpublished}{\url{https://www.nytimes.com/2019/04/22/technology/google-walkout-employees-retaliation.html}}.
\newblock \bibinfo{journal}{\emph{The New York Times}} (\bibinfo{date}{22
  April} \bibinfo{year}{2019}).
\newblock
\newblock
\shownote{Accessed: 2023-06-27}.


\bibitem[Coulton et~al\mbox{.}(2016)]%
        {coulton2016games}
\bibfield{author}{\bibinfo{person}{Paul Coulton}, \bibinfo{person}{Dan
  Burnett}, {and} \bibinfo{person}{Adrian Gradinar}.}
  \bibinfo{year}{2016}\natexlab{}.
\newblock \showarticletitle{Games as {{Speculative Design}}: {{Allowing
  Players}} to {{Consider Alternate Presents}} and {{Plausible Features}}}. In
  \bibinfo{booktitle}{\emph{Design {{Research Society Conference}} 2016}}.
\newblock
\urldef\tempurl%
\url{https://doi.org/10.21606/drs.2016.15}
\showDOI{\tempurl}


\bibitem[Cunningham et~al\mbox{.}(2023)]%
        {cunningham2023grounds}
\bibfield{author}{\bibinfo{person}{Jay Cunningham}, \bibinfo{person}{Gabrielle
  Benabdallah}, \bibinfo{person}{Daniela Rosner}, {and} \bibinfo{person}{Alex
  Taylor}.} \bibinfo{year}{2023}\natexlab{}.
\newblock \showarticletitle{On the grounds of solutionism: Ontologies of
  blackness and HCI}.
\newblock \bibinfo{journal}{\emph{ACM Transactions on Computer-Human
  Interaction}} \bibinfo{volume}{30}, \bibinfo{number}{2}
  (\bibinfo{year}{2023}), \bibinfo{pages}{1--17}.
\newblock


\bibitem[Deetz(1992)]%
        {deetz1992democracy}
\bibfield{author}{\bibinfo{person}{Stanley Deetz}.}
  \bibinfo{year}{1992}\natexlab{}.
\newblock \bibinfo{booktitle}{\emph{Democracy in an age of corporate
  colonization: Developments in communication and the politics of everyday
  life}}.
\newblock \bibinfo{publisher}{SUNY press}.
\newblock


\bibitem[Dunne and Raby(2013)]%
        {dunne2013speculative}
\bibfield{author}{\bibinfo{person}{Anthony Dunne} {and} \bibinfo{person}{Fiona
  Raby}.} \bibinfo{year}{2013}\natexlab{}.
\newblock \bibinfo{booktitle}{\emph{Speculative {Everything}: {Design},
  {Fiction}, and {Social} {Dreaming}}}.
\newblock \bibinfo{publisher}{MIT Press}.
\newblock
\showISBNx{978-0-262-01984-2}
\newblock
\shownote{Google-Books-ID: 9gQyAgAAQBAJ}.


\bibitem[Flanagan and Nissenbaum(2014)]%
        {flanagan2014values}
\bibfield{author}{\bibinfo{person}{Mary Flanagan} {and} \bibinfo{person}{Helen
  Nissenbaum}.} \bibinfo{year}{2014}\natexlab{}.
\newblock \bibinfo{booktitle}{\emph{Values at {Play} in {Digital} {Games}}}.
\newblock
\urldef\tempurl%
\url{https://doi.org/10.7551/mitpress/9016.001.0001}
\showDOI{\tempurl}


\bibitem[Gansky and McDonald(2022)]%
        {gansky2022counterfacctual}
\bibfield{author}{\bibinfo{person}{Ben Gansky} {and} \bibinfo{person}{Sean
  McDonald}.} \bibinfo{year}{2022}\natexlab{}.
\newblock \showarticletitle{CounterFAccTual: How FAccT undermines its
  organizing principles}. In \bibinfo{booktitle}{\emph{2022 ACM Conference on
  Fairness, Accountability, and Transparency}}. \bibinfo{pages}{1982--1992}.
\newblock


\bibitem[Greene et~al\mbox{.}(2019)]%
        {greene2019better}
\bibfield{author}{\bibinfo{person}{Daniel Greene}, \bibinfo{person}{Anna~Lauren
  Hoffmann}, {and} \bibinfo{person}{Luke Stark}.}
  \bibinfo{year}{2019}\natexlab{}.
\newblock \showarticletitle{Better, nicer, clearer, fairer: A critical
  assessment of the movement for ethical artificial intelligence and machine
  learning}. In \bibinfo{booktitle}{\emph{52nd Hawaii international conference
  on system sciences}}.
\newblock


\bibitem[Gururaja et~al\mbox{.}(2023)]%
        {gururaja2023build}
\bibfield{author}{\bibinfo{person}{Sireesh Gururaja}, \bibinfo{person}{Amanda
  Bertsch}, \bibinfo{person}{Clara Na}, \bibinfo{person}{David~Gray Widder},
  {and} \bibinfo{person}{Emma Strubell}.} \bibinfo{year}{2023}\natexlab{}.
\newblock \showarticletitle{To Build Our Future, We Must Know Our Past:
  Contextualizing Paradigm Shifts in Natural Language Processing}.
\newblock \bibinfo{journal}{\emph{arXiv preprint arXiv:2310.07715}}
  (\bibinfo{year}{2023}).
\newblock


\bibitem[Gusterson(1996)]%
        {gusterson1996nuclear}
\bibfield{author}{\bibinfo{person}{Hugh Gusterson}.}
  \bibinfo{year}{1996}\natexlab{}.
\newblock \bibinfo{booktitle}{\emph{Nuclear {{Rites}}: {{A Weapons Laboratory}}
  at the {{End}} of the {{Cold War}}}}.
\newblock \bibinfo{publisher}{{University of California Press}}.
\newblock
\showISBNx{978-0-520-21373-9}


\bibitem[Haraway(1991)]%
        {haraway1991situated}
\bibfield{author}{\bibinfo{person}{Donna~J Haraway}.}
  \bibinfo{year}{1991}\natexlab{}.
\newblock \showarticletitle{Situated knowledges: The science question in
  feminism and the privilege of partial perspective}.
\newblock \bibinfo{journal}{\emph{Simians, cyborgs, and women: The reinvention
  of nature}} (\bibinfo{year}{1991}), \bibinfo{pages}{183--201}.
\newblock


\bibitem[Holstein et~al\mbox{.}(2019)]%
        {holstein2019improving}
\bibfield{author}{\bibinfo{person}{Kenneth Holstein}, \bibinfo{person}{Jennifer
  Wortman~Vaughan}, \bibinfo{person}{Hal Daum{\'e}~III}, \bibinfo{person}{Miro
  Dudik}, {and} \bibinfo{person}{Hanna Wallach}.}
  \bibinfo{year}{2019}\natexlab{}.
\newblock \showarticletitle{Improving fairness in machine learning systems:
  What do industry practitioners need?}. In \bibinfo{booktitle}{\emph{2019 CHI
  conference on human factors in computing systems}}. \bibinfo{pages}{1--16}.
\newblock


\bibitem[Huber and Lewis(2010)]%
        {huber2010crossunderstanding}
\bibfield{author}{\bibinfo{person}{George~P. Huber} {and} \bibinfo{person}{Kyle
  Lewis}.} \bibinfo{year}{2010}\natexlab{}.
\newblock \showarticletitle{Cross-{Understanding}: {Implications} for {Group}
  {Cognition} and {Performance}}.
\newblock \bibinfo{journal}{\emph{The Academy of Management Review}}
  \bibinfo{volume}{35}, \bibinfo{number}{1} (\bibinfo{year}{2010}),
  \bibinfo{pages}{6--26}.
\newblock
\showISSN{0363-7425}
\urldef\tempurl%
\url{https://www.jstor.org/stable/27760038}
\showURL{%
\tempurl}
\newblock
\shownote{Publisher: Academy of Management}.


\bibitem[Humphreys(2005)]%
        {humphreys2005reframing}
\bibfield{author}{\bibinfo{person}{Lee Humphreys}.}
  \bibinfo{year}{2005}\natexlab{}.
\newblock \showarticletitle{Reframing {Social} {Groups}, {Closure}, and
  {Stabilization} in the {Social} {Construction} of {Technology}}.
\newblock \bibinfo{journal}{\emph{Social Epistemology}} \bibinfo{volume}{19},
  \bibinfo{number}{2-3} (\bibinfo{date}{Jan.} \bibinfo{year}{2005}),
  \bibinfo{pages}{231--253}.
\newblock
\showISSN{0269-1728}
\urldef\tempurl%
\url{https://doi.org/10.1080/02691720500145449}
\showDOI{\tempurl}
\newblock
\shownote{Publisher: Routledge \_eprint:
  https://doi.org/10.1080/02691720500145449}.


\bibitem[Janardhanan et~al\mbox{.}(2020)]%
        {janardhanan2020getting}
\bibfield{author}{\bibinfo{person}{Niranjan~S. Janardhanan},
  \bibinfo{person}{Kyle Lewis}, \bibinfo{person}{Rhonda~K. Reger}, {and}
  \bibinfo{person}{Cynthia~K. Stevens}.} \bibinfo{year}{2020}\natexlab{}.
\newblock \showarticletitle{Getting to {Know} {You}: {Motivating}
  {Cross}-{Understanding} for {Improved} {Team} and {Individual}
  {Performance}}.
\newblock \bibinfo{journal}{\emph{Organization Science}} \bibinfo{volume}{31},
  \bibinfo{number}{1} (\bibinfo{date}{Jan.} \bibinfo{year}{2020}),
  \bibinfo{pages}{103--118}.
\newblock
\showISSN{1047-7039}
\urldef\tempurl%
\url{https://doi.org/10.1287/orsc.2019.1324}
\showDOI{\tempurl}
\newblock
\shownote{Publisher: INFORMS}.


\bibitem[Jobin et~al\mbox{.}(2019)]%
        {jobin2019global}
\bibfield{author}{\bibinfo{person}{Anna Jobin}, \bibinfo{person}{Marcello
  Ienca}, {and} \bibinfo{person}{Effy Vayena}.}
  \bibinfo{year}{2019}\natexlab{}.
\newblock \showarticletitle{The global landscape of AI ethics guidelines}.
\newblock \bibinfo{journal}{\emph{Nature Machine Intelligence}}
  \bibinfo{volume}{1}, \bibinfo{number}{9} (\bibinfo{year}{2019}),
  \bibinfo{pages}{389--399}.
\newblock


\bibitem[Johnson(2019)]%
        {johnson2019ai}
\bibfield{author}{\bibinfo{person}{Khari Johnson}.}
  \bibinfo{year}{2019}\natexlab{}.
\newblock \showarticletitle{AI ethics is all about power}.
\newblock \bibinfo{journal}{\emph{Venture Beat}}  \bibinfo{volume}{1}
  (\bibinfo{year}{2019}).
\newblock


\bibitem[Kellogg(2009)]%
        {kellogg2009operating}
\bibfield{author}{\bibinfo{person}{Katherine~C. Kellogg}.}
  \bibinfo{year}{2009}\natexlab{}.
\newblock \showarticletitle{Operating {{Room}}: {{Relational Spaces}} and
  {{Microinstitutional Change}} in {{Surgery}}}.
\newblock \bibinfo{journal}{\emph{Amer. J. Sociology}} \bibinfo{volume}{115},
  \bibinfo{number}{3} (\bibinfo{date}{Nov.} \bibinfo{year}{2009}),
  \bibinfo{pages}{657--711}.
\newblock
\showISSN{0002-9602}
\urldef\tempurl%
\url{https://doi.org/10.1086/603535}
\showDOI{\tempurl}


\bibitem[Keyes et~al\mbox{.}(2019)]%
        {keyes2019mulching}
\bibfield{author}{\bibinfo{person}{Os Keyes}, \bibinfo{person}{Jevan Hutson},
  {and} \bibinfo{person}{Meredith Durbin}.} \bibinfo{year}{2019}\natexlab{}.
\newblock \showarticletitle{A mulching proposal: Analysing and improving an
  algorithmic system for turning the elderly into high-nutrient slurry}. In
  \bibinfo{booktitle}{\emph{Extended Abstracts of the 2019 CHI Conference on
  Human Factors in Computing Systems}}. \bibinfo{pages}{1--11}.
\newblock


\bibitem[Lee et~al\mbox{.}({[n.\,d.]})]%
        {leedon}
\bibfield{author}{\bibinfo{person}{Hao-Ping~Hank Lee}, \bibinfo{person}{Lan
  Gao}, \bibinfo{person}{Stephanie Yang}, \bibinfo{person}{Jodi Forlizzi},
  {and} \bibinfo{person}{Sauvik Das}.} \bibinfo{year}{[n.\,d.]}\natexlab{}.
\newblock \showarticletitle{“I Don’t Know If We’re Doing Good. I Don’t
  Know If We’re Doing Bad”: Investigating How Practitioners Scope,
  Motivate, and Conduct Privacy Work When Developing AI Products}.
\newblock  (\bibinfo{year}{[n.\,d.]}).
\newblock


\bibitem[Lee et~al\mbox{.}(2020)]%
        {lee2020power}
\bibfield{author}{\bibinfo{person}{Jennifer Lee}, \bibinfo{person}{Meg Young},
  \bibinfo{person}{PM Krafft}, {and} \bibinfo{person}{Michael~A Katell}.}
  \bibinfo{year}{2020}\natexlab{}.
\newblock \showarticletitle{Power and technology: Who gets to make the
  decisions?}
\newblock \bibinfo{journal}{\emph{Interactions}} \bibinfo{volume}{28},
  \bibinfo{number}{1} (\bibinfo{year}{2020}), \bibinfo{pages}{38--46}.
\newblock


\bibitem[Lincoln et~al\mbox{.}(2011)]%
        {lincoln2011paradigmatic}
\bibfield{author}{\bibinfo{person}{Yvonna~S Lincoln}, \bibinfo{person}{Susan~A
  Lynham}, {and} \bibinfo{person}{Egon~G Guba}.}
  \bibinfo{year}{2011}\natexlab{}.
\newblock \showarticletitle{Paradigmatic controversies, contradictions, and
  emerging confluences, revisited}.
\newblock \bibinfo{journal}{\emph{The Sage handbook of qualitative research}}
  \bibinfo{volume}{4} (\bibinfo{year}{2011}), \bibinfo{pages}{97--128}.
\newblock


\bibitem[Madaio et~al\mbox{.}(2022)]%
        {madaio2022assessing}
\bibfield{author}{\bibinfo{person}{Michael Madaio}, \bibinfo{person}{Lisa
  Egede}, \bibinfo{person}{Hariharan Subramonyam}, \bibinfo{person}{Jennifer
  Wortman~Vaughan}, {and} \bibinfo{person}{Hanna Wallach}.}
  \bibinfo{year}{2022}\natexlab{}.
\newblock \showarticletitle{Assessing the Fairness of AI Systems: AI
  Practitioners' Processes, Challenges, and Needs for Support}.
\newblock \bibinfo{journal}{\emph{ACM Conference on Human-Computer
  Interaction}} \bibinfo{volume}{6}, \bibinfo{number}{CSCW1}
  (\bibinfo{year}{2022}), \bibinfo{pages}{1--26}.
\newblock


\bibitem[Madaio et~al\mbox{.}(2020a)]%
        {madaio2020co}
\bibfield{author}{\bibinfo{person}{Michael~A Madaio}, \bibinfo{person}{Luke
  Stark}, \bibinfo{person}{Jennifer Wortman~Vaughan}, {and}
  \bibinfo{person}{Hanna Wallach}.} \bibinfo{year}{2020}\natexlab{a}.
\newblock \showarticletitle{Co-designing checklists to understand
  organizational challenges and opportunities around fairness in ai}. In
  \bibinfo{booktitle}{\emph{2020 CHI Conference on Human Factors in Computing
  Systems}}. \bibinfo{pages}{1--14}.
\newblock


\bibitem[Madaio et~al\mbox{.}(2020b)]%
        {madaio2020microsoft}
\bibfield{author}{\bibinfo{person}{Michael~A Madaio}, \bibinfo{person}{Luke
  Stark}, \bibinfo{person}{Jennifer Wortman~Vaughan}, {and}
  \bibinfo{person}{Hanna Wallach}.} \bibinfo{year}{2020}\natexlab{b}.
\newblock \bibinfo{title}{Microsoft AI Fairness Checklist}.
\newblock
  \bibinfo{howpublished}{https://query.prod.cms.rt.microsoft.com/cms/api/am/binary/RE4t6dA}.
\newblock


\bibitem[Mandel et~al\mbox{.}(2020)]%
        {mandel2020crowd}
\bibfield{author}{\bibinfo{person}{Travis Mandel}, \bibinfo{person}{Jahnu
  Best}, \bibinfo{person}{Randall~H. Tanaka}, \bibinfo{person}{Hiram Temple},
  \bibinfo{person}{Chansen Haili}, \bibinfo{person}{Sebastian~J. Carter},
  \bibinfo{person}{Kayla Schlechtinger}, {and} \bibinfo{person}{Roy Szeto}.}
  \bibinfo{year}{2020}\natexlab{}.
\newblock \showarticletitle{Using the Crowd to Prevent Harmful AI Behavior}.
\newblock \bibinfo{journal}{\emph{Proc. ACM Hum.-Comput. Interact.}}
  \bibinfo{volume}{4}, \bibinfo{number}{CSCW2}, Article \bibinfo{articleno}{97}
  (\bibinfo{date}{oct} \bibinfo{year}{2020}), \bibinfo{numpages}{25}~pages.
\newblock
\urldef\tempurl%
\url{https://doi.org/10.1145/3415168}
\showDOI{\tempurl}


\bibitem[Mankoff et~al\mbox{.}(2013)]%
        {mankoff2013looking}
\bibfield{author}{\bibinfo{person}{Jennifer Mankoff},
  \bibinfo{person}{Jennifer~A Rode}, {and} \bibinfo{person}{Haakon Faste}.}
  \bibinfo{year}{2013}\natexlab{}.
\newblock \showarticletitle{Looking past yesterday's tomorrow: using futures
  studies methods to extend the research horizon}.
\newblock  (\bibinfo{date}{April} \bibinfo{year}{2013}), \bibinfo{pages}{10}.
\newblock


\bibitem[Martelaro and Ju(2020)]%
        {martelaro2020could}
\bibfield{author}{\bibinfo{person}{Nikolas Martelaro} {and}
  \bibinfo{person}{Wendy Ju}.} \bibinfo{year}{2020}\natexlab{}.
\newblock \showarticletitle{What could go wrong? Exploring the downsides of
  autonomous vehicles}. In \bibinfo{booktitle}{\emph{12th International
  Conference on Automotive User Interfaces and Interactive Vehicular
  Applications}}. \bibinfo{pages}{99--101}.
\newblock


\bibitem[{Mesmer-Magnus} and DeChurch(2009)]%
        {mesmer-magnus2009information}
\bibfield{author}{\bibinfo{person}{Jessica~R. {Mesmer-Magnus}} {and}
  \bibinfo{person}{Leslie~A. DeChurch}.} \bibinfo{year}{2009}\natexlab{}.
\newblock \showarticletitle{Information Sharing and Team Performance: {{A}}
  Meta-Analysis}.
\newblock \bibinfo{journal}{\emph{Journal of Applied Psychology}}
  \bibinfo{volume}{94}, \bibinfo{number}{2} (\bibinfo{year}{2009}),
  \bibinfo{pages}{535--546}.
\newblock
\showISSN{1939-1854}
\urldef\tempurl%
\url{https://doi.org/10.1037/a0013773}
\showDOI{\tempurl}


\bibitem[Metcalf et~al\mbox{.}(2019)]%
        {metcalf2019owning}
\bibfield{author}{\bibinfo{person}{Jacob Metcalf}, \bibinfo{person}{Emanuel
  Moss}, {et~al\mbox{.}}} \bibinfo{year}{2019}\natexlab{}.
\newblock \showarticletitle{Owning ethics: Corporate logics, silicon valley,
  and the institutionalization of ethics}.
\newblock \bibinfo{journal}{\emph{Social Research: An International Quarterly}}
  \bibinfo{volume}{86}, \bibinfo{number}{2} (\bibinfo{year}{2019}),
  \bibinfo{pages}{449--476}.
\newblock


\bibitem[Metz(2023)]%
        {metz2023godfather}
\bibfield{author}{\bibinfo{person}{Cade Metz}.}
  \bibinfo{year}{2023}\natexlab{}.
\newblock \showarticletitle{‘{The} {Godfather} of {A}.{I}.’ {Leaves}
  {Google} and {Warns} of {Danger} {Ahead}}.
\newblock \bibinfo{journal}{\emph{The New York Times}} (\bibinfo{date}{May}
  \bibinfo{year}{2023}).
\newblock
\showISSN{0362-4331}
\urldef\tempurl%
\url{https://www.nytimes.com/2023/05/01/technology/ai-google-chatbot-engineer-quits-hinton.html}
\showURL{%
\tempurl}


\bibitem[Metz and Wakabayashi(2020)]%
        {metz2020google}
\bibfield{author}{\bibinfo{person}{Cade Metz} {and} \bibinfo{person}{Daisuke
  Wakabayashi}.} \bibinfo{year}{2020}\natexlab{}.
\newblock \showarticletitle{Google {{Researcher Says She Was Fired Over Paper
  Highlighting Bias}} in {{A}}.{{I}}.}
\newblock \bibinfo{journal}{\emph{The New York Times}} (\bibinfo{date}{Dec.}
  \bibinfo{year}{2020}).
\newblock
\showISSN{0362-4331}


\bibitem[Mittelstadt(2019)]%
        {mittelstadt2019principles}
\bibfield{author}{\bibinfo{person}{Brent Mittelstadt}.}
  \bibinfo{year}{2019}\natexlab{}.
\newblock \showarticletitle{Principles alone cannot guarantee ethical AI}.
\newblock \bibinfo{journal}{\emph{Nature Machine Intelligence}}
  \bibinfo{volume}{1}, \bibinfo{number}{11} (\bibinfo{year}{2019}),
  \bibinfo{pages}{501--507}.
\newblock


\bibitem[Rakova et~al\mbox{.}(2020)]%
        {rakova2020responsible}
\bibfield{author}{\bibinfo{person}{Bogdana Rakova}, \bibinfo{person}{Jingying
  Yang}, \bibinfo{person}{Henriette Cramer}, {and} \bibinfo{person}{Rumman
  Chowdhury}.} \bibinfo{year}{2020}\natexlab{}.
\newblock \showarticletitle{Where Responsible AI meets Reality: Practitioner
  Perspectives on Enablers for shifting Organizational Practices}.
\newblock \bibinfo{journal}{\emph{In 24th ACM Conference on Computer-Supported
  Cooperative Work and Social Computing}} (\bibinfo{year}{2020}).
\newblock


\bibitem[Scott(1990)]%
        {scott1990domination}
\bibfield{author}{\bibinfo{person}{James~C Scott}.}
  \bibinfo{year}{1990}\natexlab{}.
\newblock \bibinfo{booktitle}{\emph{Domination and the arts of resistance:
  Hidden transcripts}}.
\newblock \bibinfo{publisher}{Yale university press}.
\newblock


\bibitem[Skirpan et~al\mbox{.}(2022)]%
        {skirpan2022privacy}
\bibfield{author}{\bibinfo{person}{Michael Skirpan}, \bibinfo{person}{Maggie
  Oates}, \bibinfo{person}{Daragh Byrne}, \bibinfo{person}{Robert Cunningham},
  {and} \bibinfo{person}{Lorrie~Faith Cranor}.}
  \bibinfo{year}{2022}\natexlab{}.
\newblock \showarticletitle{Is a privacy crisis experienced, a privacy crisis
  avoided?}
\newblock \bibinfo{journal}{\emph{Commun. ACM}} \bibinfo{volume}{65},
  \bibinfo{number}{3} (\bibinfo{date}{March} \bibinfo{year}{2022}),
  \bibinfo{pages}{26--29}.
\newblock
\showISSN{0001-0782, 1557-7317}
\urldef\tempurl%
\url{https://doi.org/10.1145/3512325}
\showDOI{\tempurl}


\bibitem[Smith et~al\mbox{.}(2015)]%
        {smith2015metaanalysis}
\bibfield{author}{\bibinfo{person}{Shamus~P. Smith}, \bibinfo{person}{Karen
  Blackmore}, {and} \bibinfo{person}{Keith Nesbitt}.}
  \bibinfo{year}{2015}\natexlab{}.
\newblock \showarticletitle{A {{Meta-Analysis}} of {{Data Collection}} in
  {{Serious Games Research}}}.
\newblock In \bibinfo{booktitle}{\emph{Serious {{Games Analytics}}:
  {{Methodologies}} for {{Performance Measurement}}, {{Assessment}}, and
  {{Improvement}}}}, \bibfield{editor}{\bibinfo{person}{Christian~Sebastian
  Loh}, \bibinfo{person}{Yanyan Sheng}, {and} \bibinfo{person}{Dirk
  Ifenthaler}} (Eds.). \bibinfo{publisher}{{Springer International
  Publishing}}, \bibinfo{address}{{Cham}}, \bibinfo{pages}{31--55}.
\newblock
\showISBNx{978-3-319-05834-4}
\urldef\tempurl%
\url{https://doi.org/10.1007/978-3-319-05834-4_2}
\showDOI{\tempurl}


\bibitem[Strauss and Corbin(1990)]%
        {strauss1990basics}
\bibfield{author}{\bibinfo{person}{Anselm Strauss} {and}
  \bibinfo{person}{Juliet Corbin}.} \bibinfo{year}{1990}\natexlab{}.
\newblock \bibinfo{booktitle}{\emph{Basics of qualitative research}}.
\newblock \bibinfo{publisher}{Sage publications}.
\newblock


\bibitem[Su et~al\mbox{.}(2021)]%
        {su2021critical}
\bibfield{author}{\bibinfo{person}{Norman~Makoto Su}, \bibinfo{person}{Amanda
  Lazar}, {and} \bibinfo{person}{Lilly Irani}.}
  \bibinfo{year}{2021}\natexlab{}.
\newblock \showarticletitle{Critical Affects: Tech Work Emotions Amidst the
  Techlash}.
\newblock \bibinfo{journal}{\emph{ACM Conference on Human-Computer
  Interaction}} \bibinfo{volume}{5}, \bibinfo{number}{CSCW1}
  (\bibinfo{year}{2021}), \bibinfo{pages}{1--27}.
\newblock


\bibitem[Suchman(2002)]%
        {suchman2002located}
\bibfield{author}{\bibinfo{person}{Lucy Suchman}.}
  \bibinfo{year}{2002}\natexlab{}.
\newblock \showarticletitle{Located accountabilities in technology production}.
\newblock \bibinfo{journal}{\emph{Scandinavian journal of information systems}}
  \bibinfo{volume}{14}, \bibinfo{number}{2} (\bibinfo{year}{2002}),
  \bibinfo{pages}{7}.
\newblock


\bibitem[Tomlinson(2020)]%
        {tomlinson2020stasis}
\bibfield{author}{\bibinfo{person}{Elizabeth~C. Tomlinson}.}
  \bibinfo{year}{2020}\natexlab{}.
\newblock \showarticletitle{Stasis in the {Shark} {Tank}: {Persuading} an
  {Audience} of {Funders} to {Act} on {Behalf} of {Entrepreneurs}}.
\newblock \bibinfo{journal}{\emph{Journal of Business and Technical
  Communication}} (\bibinfo{date}{March} \bibinfo{year}{2020}).
\newblock
\urldef\tempurl%
\url{https://doi.org/10.1177/1050651920910219}
\showDOI{\tempurl}
\newblock
\shownote{Publisher: SAGE PublicationsSage CA: Los Angeles, CA}.


\bibitem[Varanasi and Goyal(2023)]%
        {Varanasi2023hodgepodge}
\bibfield{author}{\bibinfo{person}{Rama~Adithya Varanasi} {and}
  \bibinfo{person}{Nitesh Goyal}.} \bibinfo{year}{2023}\natexlab{}.
\newblock \showarticletitle{“It is Currently Hodgepodge”: Examining AI/ML
  Practitioners’ Challenges during Co-Production of Responsible AI Values}.
  In \bibinfo{booktitle}{\emph{Proceedings of the 2023 CHI Conference on Human
  Factors in Computing Systems}} (Hamburg, Germany) \emph{(\bibinfo{series}{CHI
  '23})}. \bibinfo{publisher}{Association for Computing Machinery},
  \bibinfo{address}{New York, NY, USA}, Article \bibinfo{articleno}{251},
  \bibinfo{numpages}{17}~pages.
\newblock
\showISBNx{9781450394215}
\urldef\tempurl%
\url{https://doi.org/10.1145/3544548.3580903}
\showDOI{\tempurl}


\bibitem[Vaughan(1996)]%
        {vaughan1996challenger}
\bibfield{author}{\bibinfo{person}{Diane Vaughan}.}
  \bibinfo{year}{1996}\natexlab{}.
\newblock \bibinfo{booktitle}{\emph{The Challenger launch decision: Risky
  technology, culture, and deviance at NASA}}.
\newblock \bibinfo{publisher}{University of Chicago press}.
\newblock


\bibitem[Veale et~al\mbox{.}(2018)]%
        {veale2018fairness}
\bibfield{author}{\bibinfo{person}{Michael Veale}, \bibinfo{person}{Max
  Van~Kleek}, {and} \bibinfo{person}{Reuben Binns}.}
  \bibinfo{year}{2018}\natexlab{}.
\newblock \showarticletitle{Fairness and accountability design needs for
  algorithmic support in high-stakes public sector decision-making}. In
  \bibinfo{booktitle}{\emph{2018 chi conference on human factors in computing
  systems}}. \bibinfo{pages}{1--14}.
\newblock


\bibitem[Weiss(1995)]%
        {weiss1995learning}
\bibfield{author}{\bibinfo{person}{Robert~S Weiss}.}
  \bibinfo{year}{1995}\natexlab{}.
\newblock \bibinfo{booktitle}{\emph{Learning from strangers: The art and method
  of qualitative interview studies}}.
\newblock \bibinfo{publisher}{Simon and Schuster}.
\newblock


\bibitem[Widder et~al\mbox{.}(2021)]%
        {widder2021trust}
\bibfield{author}{\bibinfo{person}{David~Gray Widder}, \bibinfo{person}{Laura
  Dabbish}, \bibinfo{person}{James~D Herbsleb}, \bibinfo{person}{Alexandra
  Holloway}, {and} \bibinfo{person}{Scott Davidoff}.}
  \bibinfo{year}{2021}\natexlab{}.
\newblock \showarticletitle{Trust in Collaborative Automation in High Stakes
  Software Engineering Work: A Case Study at NASA}. In
  \bibinfo{booktitle}{\emph{Proceedings of the 2021 CHI Conference on Human
  Factors in Computing Systems}}. \bibinfo{pages}{1--13}.
\newblock


\bibitem[Widder and Nafus(2023)]%
        {widder2023dislocated}
\bibfield{author}{\bibinfo{person}{David~Gray Widder} {and}
  \bibinfo{person}{Dawn Nafus}.} \bibinfo{year}{2023}\natexlab{}.
\newblock \showarticletitle{Dislocated accountabilities in the “AI supply
  chain”: Modularity and developers’ notions of responsibility}.
\newblock \bibinfo{journal}{\emph{Big Data and Society}}
  (\bibinfo{year}{2023}), \bibinfo{pages}{1--12}.
\newblock
\urldef\tempurl%
\url{https://doi.org/10.1177/20539517231177620}
\showDOI{\tempurl}


\bibitem[Widder et~al\mbox{.}(2022)]%
        {widder2022limits}
\bibfield{author}{\bibinfo{person}{David~Gray Widder}, \bibinfo{person}{Dawn
  Nafus}, \bibinfo{person}{Laura Dabbish}, {and} \bibinfo{person}{James
  Herbsleb}.} \bibinfo{year}{2022}\natexlab{}.
\newblock \showarticletitle{Limits and Possibilities for “Ethical AI” in
  Open Source: A Study of Deepfakes}. In \bibinfo{booktitle}{\emph{conference
  on fairness, accountability, and transparency}}.
\newblock


\bibitem[Widder et~al\mbox{.}(2023)]%
        {widder2023power}
\bibfield{author}{\bibinfo{person}{David~Gray Widder}, \bibinfo{person}{Derrick
  Zhen}, \bibinfo{person}{Laura Dabbish}, {and} \bibinfo{person}{James
  Herbsleb}.} \bibinfo{year}{2023}\natexlab{}.
\newblock \showarticletitle{It’s about power: What ethical concerns do
  software engineers have, and what do they (feel they can) do about them?}. In
  \bibinfo{booktitle}{\emph{Proceedings of the 2023 ACM Conference on Fairness,
  Accountability, and Transparency (ACM FAccT)}}. \bibinfo{publisher}{ACM},
  \bibinfo{address}{Chicago IL, USA}.
\newblock
\showISBNx{979-8-4007-0192-4/23/06}
\urldef\tempurl%
\url{https://doi.org/10.1145/3593013.3594012}
\showDOI{\tempurl}


\bibitem[Wilkinson(2016)]%
        {wilkinson2016brief}
\bibfield{author}{\bibinfo{person}{Phil Wilkinson}.}
  \bibinfo{year}{2016}\natexlab{}.
\newblock \showarticletitle{A brief history of serious games}. In
  \bibinfo{booktitle}{\emph{Entertainment Computing and Serious Games:
  International GI-Dagstuhl Seminar 15283, Dagstuhl Castle, Germany, July 5-10,
  2015, Revised Selected Papers}}. Springer, \bibinfo{pages}{17--41}.
\newblock


\bibitem[Wong(2021)]%
        {wong2021tactics}
\bibfield{author}{\bibinfo{person}{Richmond~Y. Wong}.}
  \bibinfo{year}{2021}\natexlab{}.
\newblock \showarticletitle{Tactics of Soft Resistance in User Experience
  Professionals' Values Work}.
\newblock \bibinfo{journal}{\emph{Proc. ACM Hum.-Comput. Interact.}}
  \bibinfo{volume}{5}, \bibinfo{number}{CSCW2}, Article
  \bibinfo{articleno}{355} (\bibinfo{date}{oct} \bibinfo{year}{2021}),
  \bibinfo{numpages}{28}~pages.
\newblock
\urldef\tempurl%
\url{https://doi.org/10.1145/3479499}
\showDOI{\tempurl}


\bibitem[Wong et~al\mbox{.}(2023)]%
        {wong2022seeing}
\bibfield{author}{\bibinfo{person}{Richmond~Y. Wong},
  \bibinfo{person}{Michael~A. Madaio}, {and} \bibinfo{person}{Nick Merrill}.}
  \bibinfo{year}{2023}\natexlab{}.
\newblock \showarticletitle{Seeing {{Like}} a {{Toolkit}}: {{How Toolkits
  Envision}} the {{Work}} of {{AI Ethics}}}.
\newblock \bibinfo{journal}{\emph{ACM Conference on Human-Computer
  Interaction}} \bibinfo{volume}{7}, \bibinfo{number}{CSCW1}
  (\bibinfo{date}{April} \bibinfo{year}{2023}), \bibinfo{pages}{1--27}.
\newblock
\urldef\tempurl%
\url{https://doi.org/10.1145/3579621}
\showDOI{\tempurl}


\bibitem[Wu et~al\mbox{.}(2022)]%
        {wu2022sustainable}
\bibfield{author}{\bibinfo{person}{Carole-Jean Wu}, \bibinfo{person}{Ramya
  Raghavendra}, \bibinfo{person}{Udit Gupta}, \bibinfo{person}{Bilge Acun},
  \bibinfo{person}{Newsha Ardalani}, \bibinfo{person}{Kiwan Maeng},
  \bibinfo{person}{Gloria Chang}, \bibinfo{person}{Fiona~Aga Behram},
  \bibinfo{person}{James Huang}, \bibinfo{person}{Charles Bai},
  \bibinfo{person}{Michael Gschwind}, \bibinfo{person}{Anurag Gupta},
  \bibinfo{person}{Myle Ott}, \bibinfo{person}{Anastasia Melnikov},
  \bibinfo{person}{Salvatore Candido}, \bibinfo{person}{David Brooks},
  \bibinfo{person}{Geeta Chauhan}, \bibinfo{person}{Benjamin Lee},
  \bibinfo{person}{Hsien-Hsin~S. Lee}, \bibinfo{person}{Bugra Akyildiz},
  \bibinfo{person}{Maximilian Balandat}, \bibinfo{person}{Joe Spisak},
  \bibinfo{person}{Ravi Jain}, \bibinfo{person}{Mike Rabbat}, {and}
  \bibinfo{person}{Kim Hazelwood}.} \bibinfo{year}{2022}\natexlab{}.
\newblock \bibinfo{title}{Sustainable {{AI}}: {{Environmental Implications}},
  {{Challenges}} and {{Opportunities}}}.
\newblock
\newblock
\showeprint[arxiv]{2111.00364}~[cs]


\bibitem[Zimmermann(2016)]%
        {zimmermann2016card}
\bibfield{author}{\bibinfo{person}{Thomas Zimmermann}.}
  \bibinfo{year}{2016}\natexlab{}.
\newblock \showarticletitle{Card-sorting: From text to themes}.
\newblock In \bibinfo{booktitle}{\emph{Perspectives on data science for
  software engineering}}. \bibinfo{publisher}{Elsevier},
  \bibinfo{pages}{137--141}.
\newblock


\end{thebibliography}
\end{document}